\documentclass[twocolumn, pra, amsmath, amssymb, superscriptaddress, longbibliography, nofootinbib, floatfix]{revtex4-2}
\usepackage[utf8]{inputenc}
\usepackage{pgfplots}
\usepackage{bbold}
\usepackage{comment}
\usepackage[colorlinks=true, linkcolor=red, allbordercolors={white}]{hyperref}
\usepackage{algorithm}
\usepackage{algpseudocode}
\usepackage{ragged2e}
\usepackage{float}
\usepackage{physics, bm}
\pgfplotsset{compat = newest}
\usepgfplotslibrary{colorbrewer}
\usepgfplotslibrary{groupplots}
\usetikzlibrary{arrows.meta}
\definecolor{color1}{HTML}{1f77b4}
\definecolor{color2}{HTML}{ff7f0e}
\definecolor{color3}{HTML}{2ca02c}
\definecolor{color4}{HTML}{d62728}
\definecolor{color5}{HTML}{9467bd}
\definecolor{color6}{HTML}{8c564b}
\definecolor{color7}{HTML}{e377c2}
\definecolor{color8}{HTML}{7f7f7f}
\definecolor{color9}{HTML}{bcbd22}
\definecolor{shadyblue}{rgb}{0.33,0.33,1}
\pgfplotsset{
  cycle list={color1\\color2\\color3\\color4\\color5\\color6\\color7\\color8\\color9\\},
}

\newcommand{\dist}{\xi}

\begin{document}

\title{Measuring Anyonic Exchange Phases Using Two-Dimensional Coherent Spectroscopy}

\author{Nico Kirchner}
\thanks{These authors contributed equally to this work.}
\affiliation{Technical University of Munich, TUM School of Natural Sciences, Physics Department, 85748 Garching, Germany}
\affiliation{Munich Center for Quantum Science and Technology (MCQST), Schellingstr. 4, 80799 M{\"u}nchen, Germany}

\author{Wonjune Choi}
\thanks{These authors contributed equally to this work.}
\affiliation{Theoretical Division, T-4 and Center for Nonlinear Studies,
Los Alamos National Laboratory, Los Alamos, New Mexico 87545, USA}

\author{Frank Pollmann}
\affiliation{Technical University of Munich, TUM School of Natural Sciences, Physics Department, 85748 Garching, Germany}
\affiliation{Munich Center for Quantum Science and Technology (MCQST), Schellingstr. 4, 80799 M{\"u}nchen, Germany}

\begin{abstract}

Identifying experimental signatures of anyons, which exhibit fractional exchange statistics, remains a central challenge in the study of two-dimensional topologically ordered systems. Previous theoretical work has shown that the threshold behavior in linear response spectroscopy can reveal the fractional exchange statistics between an anyon and its antiparticle. In this work, we extend this framework to nonlinear, two-dimensional coherent spectroscopy. We demonstrate by analyzing time-ordered four-point correlation functions that the threshold behavior of nonlinear response functions encodes the fractional statistics between general pairs of anyons that can combine to any composite topological charge. This feature in particular provides a powerful probe for unambiguously distinguishing non-Abelian anyons, which can form multiple composite charges with distinct nontrivial braid statistics. Our approach is validated using numerical simulations that are consistent with the correct fractional exchange statistics for both the Abelian anyons in the toric code and non-Abelian Ising anyons.

\end{abstract}

\maketitle

\section{Introduction}

Anyons are exotic quasiparticles that exhibit fractional exchange statistics beyond bosons and fermions~\cite{Leinaas1977, PhysRevLett.48.1144, PhysRevLett.49.957}, and may be supported in two-dimensional topologically ordered systems. One of the fundamental challenges in this field is the identification of experimental signatures that reveal the presence and nature of such anyonic excitations.
While the most intuitive and direct approaches to detecting anyonic statistics consists of interferometric setups, such as Fabry-Pérot~\cite{PhysRevB.55.2331, Nakamura2020, PhysRevX.13.011028, Werkmeister, 240319628} and Mach-Zehnder~\cite{PhysRevB.74.045319, BONDERSON20082709, PhysRevB.108.L241302} interferometers, other setups involving anyon colliders~\cite{PhysRevLett.116.156802, Lee2022, Bartolomei, PhysRevX.13.011031, PhysRevX.13.011030} or heat conductance measurements~\cite{Banerjee2017, Banerjee2018} have also been shown to contain information about anyons.

A quite different experimental setup based on linear response spectroscopy has also been shown to exhibit signatures of fractional statistics. The leading order of the linear response signal close to the threshold of exciting an anyon-antianyon pair has been found to directly correspond to the anyonic exchange statistics~\cite{PhysRevLett.118.227201, 2505.01042}.
This naturally raises the question of whether the nonlinear dynamics can be utilized to extract specific anyonic properties that are beyond the linear response setup. Indeed, nonlinear extensions to the linear response protocol have been found to feature signatures that have no analogue in linear response and may indicate the presence of anyons in the system~\cite{PhysRevB.109.075108, 2503.22792}.

In this work, we extend these studies by considering a distinct nonlinear response protocol known as two-dimensional coherent spectroscopy (2DCS)~\cite{Mukamel, Kuehn2011, Woerner_2013, Liu2025}. We find that among the four Liouville space pathways contributing to the third-order response functions, a time-ordered four-point correlation encodes the anyonic exchange statistics between general pairs of anyons near the threshold frequency.
The difference to previous works is that we consider the frequency space of the nonlinear response, while Refs.~\cite{PhysRevB.109.075108, 2503.22792} focus on the time domain. These works further consider the pump-probe spectroscopy setup, whereas we focus on the slightly different 2DCS protocol.
Our results extend the findings for linear response spectroscopy~\cite{PhysRevLett.118.227201} in the sense that the anyonic exchange statistics is accessible for anyon pairs with arbitrary composite charges / fusion channels, rather than only for anyon-antianyon pairs with trivial composite charge.
This feature is particularly interesting for non-Abelian anyons, which possess multiple fusion channels with distinct nontrivial braiding phases. We confirm our theoretical findings by comparison with numerical simulations of Abelian toric code anyons and non-Abelian Ising anyons.

The paper is structured as follows. In Section~\ref{sec:lin_res}, we briefly review the linear response result obtained in Ref.~\cite{PhysRevLett.118.227201} and discuss in detail why an extension is necessary in order to extract the most general fractional statistics. After a short introduction to 2DCS and to the specific four-point correlation function of interest in Sec.~\ref{sec:2DCS}, we discuss the underlying assumptions and the main result in Sec.~\ref{sec:anyon2DCS}. In Section~\ref{sec:results}, we present numerical simulations supporting our theoretical findings. Finally, in Section~\ref{sec:conclusion}, we conclude by summarizing our results and giving a brief outlook on possible future research directions. The analytic derivation of the main result is deferred to App.~\ref{app:chi3}.

\section{Linear Threshold Spectroscopy}
\label{sec:lin_res}

In a two-dimensional system hosting anyons, the linear response function near the threshold energy reflects the anyonic exchange statistics \cite{PhysRevLett.118.227201}. More precisely, when exciting an anyon-antianyon pair $a\overline{a}$, where $a$ denotes the anyon and $\overline{a}$ its antiparticle, the spectral function $S(\omega)$\footnote{It is assumed that there is no transfer of total momentum in the scattering process.} close to the threshold shows the power-law scaling~\cite{PhysRevLett.118.227201}
\begin{align}
	S(\omega) \propto (\omega - \Delta)^{|\alpha|} \Theta(\omega - \Delta).
\end{align}
Here, $\Delta$ denotes the energy gap of exciting the pair $a\overline{a}$ on top of the ground state, $\Theta$ the step function, and $\alpha \in (-1,1]$ encodes the exchange statistics of the anyons.
Since $S(\omega)$ is, for example, related to the dynamical structure factor measured in inelastic neutron scattering in spin systems, the statistical parameter $\alpha$ is directly accessible from experiments.
In the anyon literature~\cite{1506.05805, topologicalquantum, KITAEV20062, BONDERSON20194065, PhysRevResearch.3.033110}, $\alpha$ is described in terms of the so-called $R$-symbol
\begin{align}
	R^{a\overline{a}}_1 = e^{i\pi \alpha},
\end{align}
which represents the statistical phase $\pi\alpha$ acquired upon (counter-clockwise) exchange.

While this result enables a direct probe of anyonic exchange statistics, it only accesses the exchange phase between $a$ and its antiparticle $\overline a$, which is related to the topological spin $\theta_a$ of anyon $a$~\cite{KITAEV20062, 1506.05805, BONDERSON20194065, PhysRevResearch.3.033110}. This limitation stems from the fact that the two excited anyons must have a trivial total topological charge, as they are created from the ground state with no topological excitations.

A natural question is whether a similar spectroscopic experiment can access the full information contained in the $R$-symbols $R^{ab}_c$, which correspond to the exchange phase acquired when exchanging two anyons $a$ and $b$ whose composite topological charge is $c$.
For example, the toric code~\cite{KITAEV20032, 1506.05805, topologicalquantum} hosts the nontrivial anyons $e$, $m$, and $f$. The linear response protocol can uncover that $e$ and $m$ are self-bosons, while $f$ is a self-fermion.
The mutual semionic statistics between $e$ and $m$, however, cannot be verified because the nontrivial fusion rule $e \times m = f$ forbids the pair $e$ and $m$ from being created directly from the vacuum. Here, $\times$ denotes the fusion operator corresponding to the formation of the composite topological charge.
 Another interesting case is non-Abelian anyons, which are characterized by having multiple possible fusion outcomes. A paradigmatic example is the Ising anyon $\sigma$~\cite{KITAEV20062, 1506.05805, topologicalquantum}, which can fuse to either a fermionic ($\psi$) or a trivial / bosonic ($1$) total charge: $\sigma \times \sigma = 1 + \psi$.
 The corresponding $R$-symbols, $R^{\sigma\sigma}_1$ and $R^{\sigma\sigma}_{\psi}$, describe the exchange phases of two Ising anyons when their total charge is 1 or $\psi$, respectively. The linear threshold spectroscopy can only measure $R^{\sigma\sigma}_1$.

In the abstract description of anyon models, the $R$-symbols are not gauge-invariant objects and are therefore not directly observable since only gauge-invariant quantities such as topological spins $\theta_a$ can be measured. The $R$-symbols and the topological spins are however related by the ribbon property~\cite{KITAEV20062, 1506.05805, topologicalquantum, BONDERSON20194065, PhysRevResearch.3.033110}
\begin{align}
	R^{ab}_cR^{ba}_c = \frac{\theta_c}{\theta_a \theta_b}.
	\label{eq:Ribbon}
\end{align}
If we choose a gauge such that $R^{ab}_c = R^{ba}_c$~\cite{PhysRevResearch.3.033110}, the $R$-symbols correspond to a ratio of gauge-invariant topological spins, which may therefore be experimentally observed. Throughout this work, we refer to the $R$-symbols in this gauge when talking about measuring the exchange statistics. Equation~\eqref{eq:Ribbon} also tells us (up to a factor of $-1$) how $R^{a\overline{a}}_1$ is related to $\theta_a$ since $\theta_1=1$ and $\theta_a=\theta_{\overline{a}}$ ~\cite{KITAEV20062, BONDERSON20194065, PhysRevResearch.3.033110}.
Although one might expect linear spectroscopy to be sufficient, as all exchange phases can in principle be inferred from the topological spins, this is not the case in practice. For systems with multiple anyon species, the response from anyons with smaller excitation gaps masks the threshold behavior of those with larger gaps. Hence, an alternative method is necessary to directly probe the general exchange phases $R_c^{ab}$.

\section{Two-Dimensional Coherent Spectroscopy}
\label{sec:2DCS}

\begin{figure}[t]
    \centering
    \includegraphics{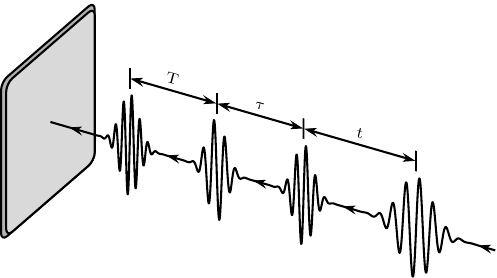}
    \caption{Schematics of the two-dimensional coherent spectroscopy setup consisting of three pulses. The time delays between the first and second, second and third, and third pulse and measurement are denoted by $T$, $\tau$, $t$, respectively. For a fixed waiting time $T$, a two-dimensional Fouier transform is performed with respect to $\tau$ and $t$.}
   \label{fig:2DCS}
\end{figure}

Two-dimensional coherent spectroscopy (2DCS)~\cite{Mukamel, Kuehn2011, Woerner_2013, Liu2025} is a particularly useful experimental technique for probing fractional excitations such as anyons since, unlike in linear spectroscopy, the associated signals can be distinguished from broadening arising from disorder and thermal effects~\cite{PhysRevLett.122.257401, Liu2025}. The 2DCS setup we consider here is schematically depicted in Fig.~\ref{fig:2DCS}. It consists of three consecutive pulses that have relative time delays of $T$ and $\tau$; the nonlinear response is measured with a delay of $t$ after the third pulse. Assuming that the pulses can be approximated by $\delta$-functions, the above protocol can be expressed as
\begin{align}
	A_0\delta(s) + A_T\delta(s-T) + A_{T+\tau}\delta(s-T-\tau),
\end{align}
where $A_{0,T,T+\tau}$ denote the amplitudes of the pulses and $s$ the time. The nonlinear response is then measured at $s=T+\tau+t$, which gives access to the nonlinear susceptibility $\chi^{(3)}(T,\tau,t)$. Its Fourier transform $\chi^{(3)}(T,\omega_{\tau},\omega_t)$ then contains information about the excitations in the system and is the main interest in this work.

In general, the susceptibility $\chi^{(3)}(T,\tau,t)$ can be computed as expectation value of nested commutators of local observables acting on the density matrix of the system~\cite{PhysRevLett.124.117205}. The various contributions of these nested commutators, also known as Liouville space pathways, correspond to signals in different regions of the $\omega_{\tau}$-$\omega_t$ plane, which allows us to isolate a specific pathway. To probe general anyon exchange phases, we focus on the time-ordered pathway, in which all interactions act on the ket side of the density matrix. Then, the anyon content evolves in time as
\begin{align}
	|0\rangle \langle 0| \xrightarrow{\, \,\,\, \,}  |{a\overline{a}} \rangle \langle 0| \xrightarrow{\,  T \,}  |{abc}\rangle \langle 0| \xrightarrow{\, \tau \,}  |{a\overline{a}} \rangle \langle 0| \xrightarrow{\, t \,}  |0\rangle\langle 0|,
	\label{eq:pathway}
\end{align}
where the arrows indicate the change of the quantum state induced by the successive pulses. The state $|0\rangle$ corresponds to the ground state without anyonic excitations, $|{a\overline{a}} \rangle$ denotes a state containing a single $a\overline{a}$ pair, and $|{abc} \rangle$ features a three-anyon state formed when the antianyon $\overline a$ splits into $b$ and $c$. Intuitively, the oscillation of the three-anyon state $|{abc}\rangle$ with respect to the ground state $|0\rangle$ should contain information about the anyonic exchange phase corresponding to $R^{bc}_{\overline{a}}$, which is confirmed below.
Other Liouville space pathways for $\chi^{(3)}$ involve out-of-time-ordered correlations, which do not directly reveal this exchange phase because they include excitations on the bra side of the density matrix. We therefore focus on the contribution to $\chi^{(3)}(T,\omega_{\tau},\omega_t)$ arising from processes of the form Eq.~\eqref{eq:pathway}, denoted by $\mathcal{C}(T,\omega_{\tau},\omega_t)$.


\section{Application of 2DCS to Anyonic Systems}
\label{sec:anyon2DCS}

Having introduced the concept of 2DCS and the relevant contribution to study, we now apply it to topologically order systems, assuming the conditions discussed below to enable analytic calculations.

\paragraph*{System:}
We consider a topologically ordered phase supporting distinct anyonic charges $\lbrace a,b,c,\ldots \rbrace$ with excitation gaps $\lbrace \Delta_a,\ldots \rbrace$ and effective masses $\lbrace m_a,\ldots \rbrace$. The spectral gaps $\Delta_{a\overline{a}} = \Delta_a + \Delta_{\overline{a}}$ for creating pairs of anyons are taken to be distinct for each anyon pair $a\overline{a}$, preventing overlap of the threshold behavior of $\mathcal{C}$ with other contributions to $\chi^{(3)}$. The same assumption applies to the triplet gaps $\Delta_{abc}=\Delta_a+\Delta_b+\Delta_c$. The energy of each anyon is approximated by a parabolic dispersion, $E_a = \mathbf{k}^2/(2m_a)$, where $\mathbf{k}$ is the momentum of the anyon. Finally, we assume that there are no non-anyonic excitations at low energies, ensuring that the nonlinear response close to the threshold is solely determined by the anyonic degrees of freedom.

\paragraph*{Drive:}
The external drive is modeled as a sequence of $\delta$-function pulses,
\begin{align}
	A(s) = A_0\delta(s) + A_T\delta(s-T) + A_{T+\tau}\delta(s-T-\tau),
\end{align}
where $s$ denotes the time (cf. Fig.~\ref{fig:2DCS}). The pulses are assumed to act on the system as
\begin{align}
H_{\text{drive}} = - A(s)
	\int \mathrm{d}\mathbf{R}\int \mathrm{d}\phi \sum_m c_m e^{im\phi} \mathcal{O}(\mathbf{R},\dist,\phi),
	\label{eq:drive}
\end{align}
where the first integral runs over all positions $\mathbf{R}$ in the system. The operator $\mathcal{O}(\mathbf{R},\dist,\phi)$ either creates an anyon pair from the vacuum ($1\rightarrow a\overline{a}$) or splits an existing anyon into two other anyons ($\overline{a} \rightarrow bc$) at positions $\mathbf{R} \pm \bm{\dist}$. The parameter $\dist$ is the characteristic distance between the created anyon pair, and $\phi$ specifies their relative orientation. The coefficients $c_m$ ($m\in \mathbb{Z}$) encode the contribution of each relative angular momentum channel involved in the anyon creation process.
The local operator $\mathcal{O}$ and $c_m$ are assumed to be time independent and spatially uniform since we consider the long wave length limit.
The pulses are further assumed to not transfer any total momentum to the system, which corresponds to illumination orthogonal to the surface of the system. The incident pulses are assumed to act only by creating, annihilating, splitting, or fusing anyons, rather than by merely increasing the kinetic energy of existing anyons without changing their particle number.
We also introduce the matrix elements $M_1^{a\overline{a}} \neq 0$ and $M_{\overline{a}}^{bc} \neq 0$, defined as
\begin{align}
	\langle a\overline{a}| \mathcal{O}(\mathbf{R},\dist,\phi) |0\rangle &=M_1^{a\overline{a}}\psi_{a\overline{a}}^{(l)}(\mathbf{R}, \dist, \phi),\\
	\langle abc| \mathcal{O}(\mathbf{R},\dist,\phi) |a\overline{a}\rangle &= M_{\overline{a}}^{bc} \psi_a(\mathbf{r}_a) \psi_{bc}^{(l)}(\mathbf{R}, \dist, \phi) \delta(\mathbf{R}-\mathbf{r}_{\overline{a}}),
\end{align}
respectively, where $\psi_{a\overline{a}}^{(l)}$ and $\psi_a\psi_{bc}^{(l)}$ denote the real space representations of the two- and three-anyon wave functions introduced below.
These matrix elements depend on the microscopic details of the system and are treated as phenomenological parameters for the purpose of this work.

\paragraph*{Anyonic wave functions:}
The two-anyon wave function for an anyon-antianyon pair $a\overline{a}$ is given by
\begin{align}
	\psi_{a\overline{a}}^{(l)}(\mathbf{R}, r, \phi) \propto \sqrt{\frac{k}{L^3}} J_{|l-\alpha|}(kr)e^{i(l-\alpha)\phi}e^{i\mathbf{K}\cdot \mathbf{R}},
	\label{eq:2anyonWF}
\end{align}
where $J_{\mu}$ denotes the Bessel function of the first kind, $L$ is the linear system size, $\mathbf{K}$ and $\mathbf{R}$ are the center of mass momentum and position, $k$ and $r$ describe the relative momentum and distance, $\phi$ is the relative orientation, and $l-\alpha$ is the anyonic angular momentum~\cite{PhysRevLett.118.227201}. The statistical parameter $\alpha$ characterizes the exchange phase of the $a\overline{a}$ pair, and $l\in 2\mathbb{Z}$. The corresponding energy of two anyons is~\cite{PhysRevLett.118.227201}
\begin{align}
	E_{a\overline{a}} = \frac{\mathbf{K}^2}{2m_{\text{CM}}} + \frac{k^2}{2\mu_{a\overline{a}}},
	\label{eq:2anyonE}
\end{align}
with $m_{\text{CM}} = m_a + m_{\overline{a}}$ and $\mu_{a\overline{a}} = m_a m_{\overline{a}} / (m_a + m_{\overline{a}})$.

In our calculations we approximate the three-anyon wave function for anyons $a$, $b$ and $c$, that originates from the ground state via pair creation of $a\overline{a}$ and subsequent splitting of $\overline{a}$ into the pair $bc$, as product state of a plane wave $\ket{a}$ describing $a$ and a two-anyon wave function $\ket{{bc}}$ describing $bc$,
\begin{align}
	|{abc}\rangle \approx |a\rangle \otimes |{bc}\rangle.
	\label{eq:3anyonWF}
\end{align}
The two-anyon wavefunction $\psi_{bc}^{(l)}(\vb R, r,\phi) = \braket{\vb R, r,\phi}{{bc}}$ has the form of Eq.~\eqref{eq:2anyonWF}, with the statistical parameter $\beta$ characterizing the exchange phase $R^{bc}_{\overline{a}} = e^{i\pi \beta}$.

The energy associated with $|{abc}\rangle$ is (cf. assumption regarding the energy above)
\begin{align}
	E_{abc} = E_a + E_{bc},
	\label{eq:3anyonE}
\end{align}
where $E_a = \mathbf{k}^2/(2m_a)$ is the single particle kinetic energy and $E_{bc}$ the two-anyon energy Eq.~\eqref{eq:2anyonE}.

The approximation in Eq.~\eqref{eq:3anyonWF} is expected to be valid when the splitting process $\overline{a}\rightarrow bc$ takes place while $a$ and $\overline{a}$ are spatially well-separated. Because the drive in Eq.~\eqref{eq:drive} can transfer any allowed angular momentum, we include all angular momentum channels for both the two- and three-anyon wave functions. Note that the associated energies Eqs.~\eqref{eq:2anyonE} and \eqref{eq:3anyonE} do not depend on the angular momenta. The effect of higher angular momenta on the threshold behavior is nevertheless suppressed (cf. App.~\ref{app:chi3}), as one would intuitively expect.

With the above assumptions, the contribution of the relevant pathway in Eq.~\eqref{eq:pathway} in the limit of small frequencies, that is, at the threshold, can be computed to be
\begin{align}
	\begin{split}
	|\mathcal{C}(T,\omega_{\tau},\omega_{t})| &\propto  \Theta(T) \sum_{a,b,c} m(a,b,c)  |M_1^{a\overline{a}}|^2  |M_{\overline{a}}^{bc}|^2  \\
	&\quad \times (\dist^2 \Omega_{t})^{|\alpha|} (\dist^2 \Omega_{\tau} - \dist\dist' \Omega_{t})^{|\beta|}\\
	&\quad \times \Theta(\Omega_{t}) \Theta\left( \dist\Omega_{\tau} - \dist'\Omega_{t} \right)  ,
	\end{split}
	\label{eq:chi3_threshold}	
\end{align}
where in addition to the quantities already defined above, $\Theta$ denotes the step function, and we introduced
\begin{align}
	\Omega_{\tau} &= \omega_{\tau} - \Delta_{abc},\qquad
	\Omega_{t} = \omega_{t} - \Delta_{a\overline{a}},\\
	\dist' &= \dist \frac{\mu_{a\overline{a}}}{m_a} \frac{m_a+m_b+m_c}{m_b+m_c},
\end{align}
and $m(a,b,c)$, which is a frequency independent quantity that depends on the anyons $a$, $b$ and $c$. The summation over all anyon species in Eq.~\eqref{eq:chi3_threshold} assures that all possible creation and splitting processes are considered. Combinations of $a$, $b$ and $c$ that are disallowed due to constraints on the anyons themselves (fusion), or on the system and drive, have $M_1^{a\overline{a}}=0$ and / or $M_{\overline{a}}^{bc}=0$. For each anyon triplet $a,b,c$, the statistical parameters $\alpha$ and $\beta$ correspond to the exchange phases $R^{a\overline{a}}_1$ and $R^{ab}_c$ via
\begin{align}
	R^{a\overline{a}}_1 = e^{i\pi \alpha}, \qquad R^{bc}_{\overline{a}} = e^{i\pi \beta},
\end{align}
with $\alpha,\beta \in (-1, 1]$. Note that the considered contribution $\mathcal{C}$ is nonzero only in the first quadrant of the $\omega_{\tau}$-$\omega_t$ plane. The analytic derivation of Eq.~\eqref{eq:chi3_threshold} can be found in App.~\ref{app:chi3}.

It can be seen from Eq.~\eqref{eq:chi3_threshold} that the threshold behavior of $|\mathcal{C}(T,\omega_{\tau},\omega_{t}=\Delta_{a\overline{a}})|$ in $\Omega_{\tau}$ with the exponent $|\beta|$ can be used to directly probe the statistcial exchange parameter $\beta$ associated with $R^{bc}_{\overline{a}}$. Such a direct probe of the exchange phase between two anyons with nontrivial composite charge is beyond linear response threshold spectroscopy (cf. the discussion in Sec.~\ref{sec:lin_res}) and is thus the main result of this work.

In addition, the threshold behavior along the direction $\dist\Omega_{\tau} = \dist'\Omega_{t}$ with the exponent $|\alpha|$ can be used as an independent probe to confirm the results for the statistical parameter $\alpha$ from linear response threshold spectroscopy. Due to the presence of the factor $\Theta ( \dist\Omega_{\tau} - \dist'\Omega_{t} )$ in Eq.~\eqref{eq:chi3_threshold}, this direction coincides with the direction of the nontrivial boundary along which $|\mathcal{C}(T,\omega_{\tau},\omega_{t})|$ becomes nonzero\footnote{There are some broadening effects, such that in order to unambiguously identify this direction, the broadening parameters must be chosen to be sufficiently small, cf. Sec.~\ref{sec:results}.}
. This direction can thus be found directly in the frequency diagram \emph{without} relying on the knowledge of additional quantities, such as the effective masses of the anyons.

Note that the threshold behavior of $|\mathcal{C}(T,\omega_{\tau}=\Delta_{abc},\omega_{t})|$ along $\Omega_{t}$ is not suitable to extract the exponent $|\alpha| + |\beta|$. Due to $\Omega_{\tau}=0$ and the factor $\Theta(\Omega_{t}) \Theta\left( \dist\Omega_{\tau} - \dist'\Omega_{t} \right)$ in Eq.~\eqref{eq:chi3_threshold}, there is no signal in $\Omega_{t}$ direction (except for potential broadening effects that we do not study).

In general, it is possible that the threshold behavior associated with an anyon triplet $abc$ is concealed by the response of another triplet $a'b'c'$ if $\Delta_{a'\overline{a'}} < \Delta_{a\overline{a}}$ and $\Delta_{a'b'c'} < \Delta_{abc}$. It is thus important to realize that in some setups, only the threshold behaviors of the anyon triplet with the smallest spectral gaps are accessible. This is analoguous to the linear response case, where we can only expect to measure the correct threshold behavior of the anyon with the smallest spectral gap. For cases with $\Delta_{a'\overline{a'}} < \Delta_{a\overline{a}}$ and $\Delta_{a'b'c'} > \Delta_{abc}$ however, we expect that the threshold spectra associated with both $abc$ and $a'b'c'$ are accessible. The reason for this is that in the former case ($\Delta_{a'\overline{a'}} < \Delta_{a\overline{a}}$, $\Delta_{a'b'c'} < \Delta_{abc}$), the threshold point at $(\omega_{\tau},\omega_{t})=(\Delta_{abc}, \Delta_{a\overline{a}})$ may lie within the region in frequency space to which the triplet $a'b'c'$ contributes. This would then result in the response associated with $a'b'c'$ concealing the threshold behavior of $abc$. For the latter case ($\Delta_{a'\overline{a'}} < \Delta_{a\overline{a}}$, $\Delta_{a'b'c'} > \Delta_{abc}$) the threshold point of each triplet does not lie within the region affected by the respective other triplet, such that the threshold behavior associated with both $abc$ and $a'b'c'$ are accessible.


\section{Numerical Results}
\label{sec:results}

For the numerical simulations, the method of exact diagonalization is used, where the Hamiltonian describing the anyons on the lattice is an anyonic tight-binding Hamiltonian~\cite{PhysRevB.43.2661, PhysRevB.43.10761, PhysRevB.107.195129} that directly incorporates the effects of the nontrivial exchange statistics. For all simulations, the lattice is chosen to be a $L_x\times L_y$ square lattice with periodic boundary conditions, which allows for the utilization of momentum states in order to reach larger system sizes. Further, nontrivial fluxes $\Phi_x,\Phi_y\in \lbrace i\phi_0/5: i\in \mathbb{Z}_5 \rbrace$ with the flux quantum $\phi_0$ are added along the two directions and averaged over in order to mitigate finite size effects.

In the following, we focus on the pathway discussed in Sec.~\ref{sec:2DCS} and only consider a single anyon pair $a\overline{a}$ and triplet $abc$. This can be justified by assuming that $a\overline{a}$ ($abc$) is the pair (triplet) with the smallest spectral gap, such that the signals associated with other anyons cannot influence or mask their threshold behavior. The relevant contribution to $\mathcal{C}(T,\omega_{\tau},\omega_{t})$ can thus be expressed as
\begin{widetext}
\begin{align}
	\mathcal{C}(T,\omega_{\tau},\omega_{t}) &= \frac{i^3}{L_xL_y} \int_{0}^{\infty}\mathrm{d}t \int_{0}^{\infty}\mathrm{d}\tau\,\, e^{i\omega_{t}t}e^{i\omega_{\tau}\tau}  \langle 0| \mathcal{O}_{0\rightarrow a\overline{a}}^{\dagger}  e^{-i\mathcal{H}_{a\overline{a}}t}   \mathcal{O}_{\overline{a}\rightarrow bc}^{\dagger}    e^{-i\mathcal{H}_{abc}\tau}   \mathcal{O}_{\overline{a}\rightarrow bc}  e^{-i\mathcal{H}_{a\overline{a}}T}   \mathcal{O}_{0\rightarrow a\overline{a}} |0 \rangle \label{eq:sim1}\\
	&= \frac{i}{L_xL_y} \langle 0| \mathcal{O}_{0\rightarrow a\overline{a}}^{\dagger}  
	\frac{1}{\omega_{t} -\mathcal{H}_{a\overline{a}} +i\epsilon_{t}}  \mathcal{O}_{\overline{a}\rightarrow bc}^{\dagger}  \frac{1}{\omega_{\tau} -\mathcal{H}_{abc} +i\epsilon_{\tau}}   \mathcal{O}_{\overline{a}\rightarrow bc}  e^{-i\mathcal{H}_{a\overline{a}}T}   \mathcal{O}_{0\rightarrow a\overline{a}} |0 \rangle,
	\label{eq:sim2}
\end{align}
\end{widetext}
where $|0\rangle$ denotes the vacuum state in which anyons are absent. The operators $\mathcal{O}_{0\rightarrow a\overline{a}}$ and $\mathcal{O}_{\overline{a}\rightarrow bc}$ are lattice realizations of the operator $\mathcal{O}(\mathbf{R},\dist,\phi)$ introduced above, with $\dist$ chosen to be the lattice spacing. $\mathcal{O}_{0\rightarrow a\overline{a}}$ excites the anyon-antianyon pair $a\overline{a}$ on nearest-neighboring sites on top of the vacuum $|0\rangle$ in all possible configurations as superposition. Similarly, $\mathcal{O}_{\overline{a}\rightarrow bc}$ splits the anyon $\overline{a}$ into $b$ and $c$ such that $b$ and $c$ occupy nearest-neighboring sites located in opposite directions. That is, for $m_b=m_c$, the center of mass of $b$ and $c$ is located precisely on the site previously occupied by $\overline{a}$. The Hamiltonians $\mathcal{H}_{a\overline{a}}$ and $\mathcal{H}_{abc}$ denote the anyonic tight-binding Hamiltonians associated with the anyon pair $a\overline{a}$ and the anyon triplet $abc$, respectively. Note that broadening parameters $\epsilon_{\tau},\epsilon_{t} > 0$ are introduced in order to ensure convergence of the integrals in Eq.~\eqref{eq:sim1} by letting $\omega_{\tau} \longrightarrow \omega_{\tau} + i\epsilon_{\tau}$ and $\omega_{t} \longrightarrow \omega_{t} + i\epsilon_{t}$. While the limit $\epsilon_{\tau},\epsilon_{t}\longrightarrow 0^+$ is usually taken in analytic calculations, we need to choose appropriate finite values for numerical simulations. In particular, it is important to use distinct values for $\epsilon_{\tau}$ and $\epsilon_{t}$ since the different density of states associated with the Hamiltonians $\mathcal{H}_{a\overline{a}}$ and $\mathcal{H}_{abc}$ lead to broadening effects of unequal strength.

\begin{figure*}[t]
\centering
\begin{minipage}[t]{0.483\textwidth}
  \centering
  \includegraphics{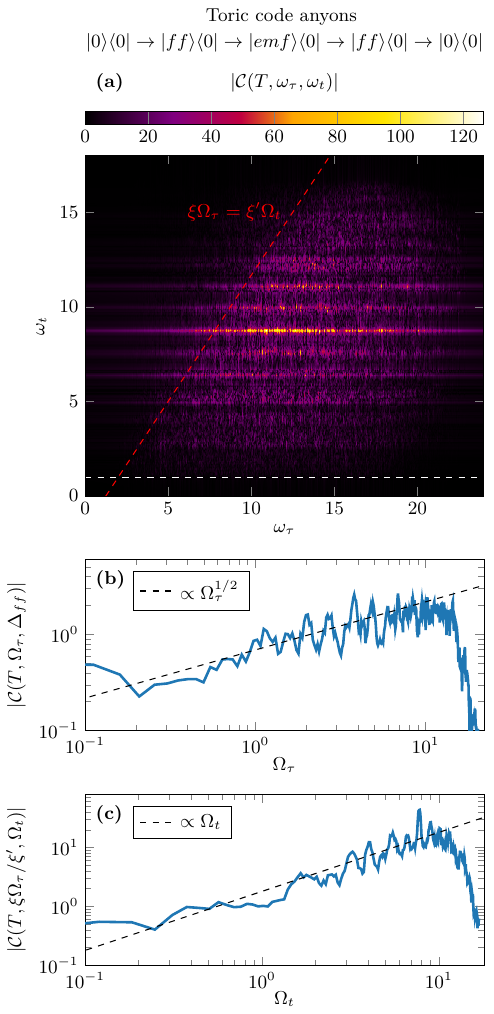}
  \caption{Simulation results for toric code anyons on a $10\times 10$ lattice with $T=7$, $\epsilon_{\tau}=5\cdot 10^{-4}$, $\epsilon_{t}=5\cdot 10^{-2}$, $m_{e}=m_{m}=m_f$, $\Delta_{ff}=1$, and $\Delta_{emf}=2$ for the process in Eq.~\eqref{eq:TCpathway}. \textbf{(a)} Up to broadening effects, $\mathcal{C}$ is only nonzero with the region bounded by the white and red dashed lines corresponding to $\Omega_{t}=0 $ and $ \dist\Omega_{\tau} = \dist'\Omega_{t}$, respectively. \textbf{(b)} The threshold behavior in $\Omega_{\tau}$ along the white dashed line is consistent with the expected exponent of $\beta=1/2$ associated with mutual semions. \textbf{(c)} The threshold behavior along the red dashed line $\dist \Omega_{\tau}=\dist' \Omega_t$ is consistent with the expected exponent of $\alpha=1$ for fermions.}
   \label{fig:chi3_TC}
\end{minipage}\hfill
\begin{minipage}[t]{0.483\textwidth}
  \centering
  \includegraphics{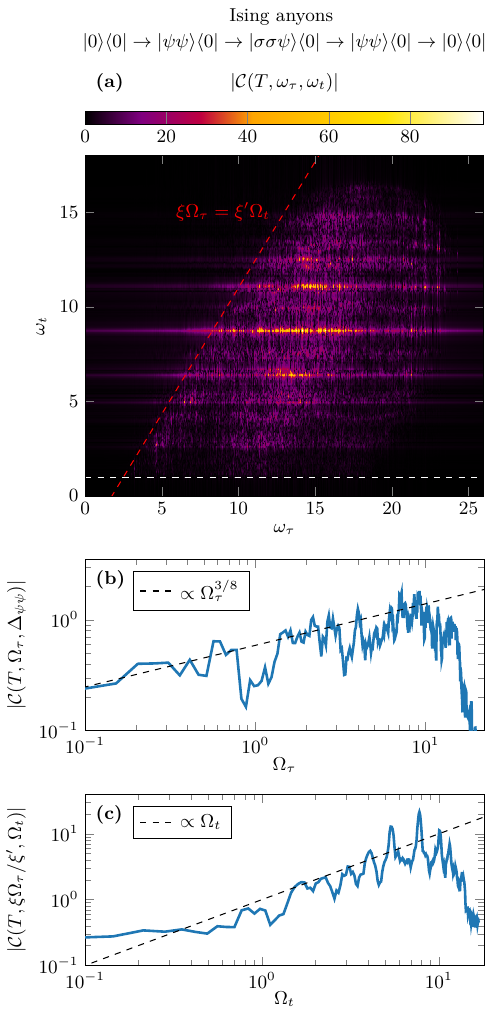}
  \caption{Simulation results for Ising anyons on a $10\times 10$ lattice with $T=7$, $\epsilon_{\tau}=5\cdot 10^{-4}$, $\epsilon_{t}=5\cdot 10^{-2}$, $m_{\sigma}=m_{\psi}$, $\Delta_{\psi\psi}=1$, and $\Delta_{\sigma\sigma\psi}=2.5$ for the process in Eq.~\eqref{eq:Isingpathway}. \textbf{(a)} Up to broadening effects, $\mathcal{C}$ is only nonzero with the region bounded by the white and red dashed lines corresponding to $\Omega_{t}=0 $ and $ \dist\Omega_{\tau} = \dist'\Omega_{t}$, respectively. \textbf{(b)} The threshold behavior in $\Omega_{\tau}$ along the white dashed line is consistent with the expected exponent of $\beta=3/8$ associated with the exchange phase $R^{\sigma\sigma}_{\psi}$. \textbf{(c)} The threshold behavior along the red dashed line $\dist \Omega_{\tau}=\dist' \Omega_t$ is consistent with the expected exponent of $\alpha=1$ for fermions.}
   \label{fig:chi3_Ising}
\end{minipage}
\end{figure*}

As a first example, we consider the Abelian anyons of the toric code~\cite{KITAEV20032, 1506.05805, topologicalquantum}, which are denoted by $e$, $m$, and $f$. Both $e$ and $m$ and self-bosons, while $f$ is a self-fermion. Since each of these anyons is its own antiparticle, this implies that using the linear response threshold spectroscopy~\cite{PhysRevLett.118.227201} described in Sec.~\ref{sec:lin_res}, only bosonic or fermionic exchange statistics can be observed. When considering our 2DCS setup on the other hand, we are able to probe the nontrivial exchange statistics between $e$ and $m$ by studying the process
\begin{align}
	|0\rangle \langle 0| \rightarrow  |ff \rangle \langle 0| \xrightarrow{  T }  |{emf}\rangle \langle 0| \xrightarrow{\,\tau\,}  |{ff} \rangle \langle 0| \xrightarrow{\, t\,}  |0\rangle\langle 0| ,
	\label{eq:TCpathway}
\end{align}
for which we expect an exponent of $\beta=1/2$ for the threshold behavior Eq.~\eqref{eq:chi3_threshold} since $e$ and $m$ are mutual semions. Figure~\ref{fig:chi3_TC} depicts the simulation results for the above process on a $10\times 10$ lattice for $T=7$, $\epsilon_{\tau}=5\cdot 10^{-4}$, $\epsilon_{t}=5\cdot 10^{-2}$, and identical effective masses $m_{e}=m_{m}=m_f$, such that $\dist'/\dist = 3/4$. The two-dimensional response in Fig.~\ref{fig:chi3_TC}\textbf{a} shows that $\mathcal{C}$ only becomes nonzero up to broadening effects if the conditions $\Theta(\Omega_{t}) $ (white dashed line) and $\Theta\left( \dist\Omega_{\tau} - \dist'\Omega_{t} \right)$ (red dashed line), cf. Eq.~\eqref{eq:chi3_threshold}, are fulfilled. The response $\mathcal{C}$, and therefore also $\chi^{(3)}$, can thus be used to determine the direction $\dist \Omega_{\tau}=\dist' \Omega_t$, along which the threshold behavior is expected to follow the exponent $\alpha=1$ corresponding to fermionic statistics. Figure~\ref{fig:chi3_TC}\textbf{b} and \textbf{c} show $\mathcal{C}$ along the dashed white and red line, respectively, where the each data point has been averaged locally with its nearest neighbors in order to mitigate the oscillations occuring due to finite size effects and the choice of $\epsilon_{\tau}, \epsilon_{t}$. In both cases, it is observed that the threshold behavior is consistent with the theoretical predictions $\beta=1/2$ and $\alpha=1$.

Let us now consider the non-Abelian Ising anyons~\cite{KITAEV20062, 1506.05805, topologicalquantum} denoted by $\sigma$ as a second example. Two Ising anyons can combine to a trivial ($1$) or fermionic ($\psi$) total topological charge, meaning that Ising anyons have different exchange phases depending on this very composite charge, that is, we have both $R^{\sigma\sigma}_1$ and $R^{\sigma\sigma}_{\psi}$. While the phase $R^{\sigma\sigma}_1$ can in principle be probed in the linear response threshold spectroscopy~\cite{PhysRevLett.118.227201}, $R^{\sigma\sigma}_{\psi}$ cannot. Within our 2DCS setup, we can find the process
\begin{align}
	|0\rangle \langle 0| \rightarrow  |\psi\psi \rangle \langle 0| \xrightarrow{  T }  |{\sigma\sigma\psi}\rangle \langle 0| \xrightarrow{\tau}  |{\psi\psi} \rangle \langle 0| \xrightarrow{ t}  |0\rangle\langle 0| ,
	\label{eq:Isingpathway}
\end{align}
which does allow to probe the fractional exchange phase $R^{\sigma\sigma}_{\psi}=e^{3i\pi /8}$~\cite{KITAEV20062, topologicalquantum} that corresponds to a threshold exponent of $\beta=3/8$. Figure~\ref{fig:chi3_Ising} depicts the simulation results for the above process on a $10\times 10$ lattice for $T=7$, $\epsilon_{\tau}=5\cdot 10^{-4}$, $\epsilon_{t}=5\cdot 10^{-2}$, and identical effective masses $m_{\sigma}=m_{\psi}$ ($\dist'/\dist = 3/4$). Similar to the toric code results, Fig.~\ref{fig:chi3_Ising}\textbf{a} shows that $\mathcal{C}$ is nonzero up to broadening effects only within the region bounded by the two dashed lines that correspond to the factors $\Theta(\Omega_{t})$ and $\Theta\left( \dist\Omega_{\tau} - \dist'\Omega_{t} \right)$ in $\mathcal{C}$. The threshold behavior of $\mathcal{C}$ in $\Omega_{\tau}$ (along the white dashed line) and in $\dist \Omega_{\tau}=\dist' \Omega_t$ direction (along the red dashed line) is depicted in Fig.~\ref{fig:chi3_Ising}\textbf{b} and \textbf{c}, respectively. Similar to the toric code, the data has been averaged in order to mitigate the oscillations. The results are found to be consistent with the theoretical expectations of $\beta=3/8$ for $R^{\sigma\sigma}_{\psi}$ and $\alpha=1$ for the fermionic statistics. Note however that the deviations from the expected behavior in particular in Fig.~\ref{fig:chi3_Ising}\textbf{b} are larger than for the toric code anyons. We attribute these deviations to finite size effects since consistency with the predictions is still established.

Overall, we have thus demonstrated for the toric code and Ising anyons that the analytical prediction of the threshold behavior in Eq.~\eqref{eq:chi3_threshold} agrees with the simulation results. The discussed 2DCS setup is thus expected to be a useful tool for extracting anyonic exchange phases.



\section{Conclusion}
\label{sec:conclusion}

We studied how nonlinear response, measurable through two-dimensional coherent spectroscopy, can reveal the braiding statistics of anyonic excitations in topologically ordered systems. From the time-ordered four-point correlation contributing to the third-order response function, we showed that the full information contained in the $R$-symbols of non-Abelian anyons, as well as the mutual statistics of Abelian anyons with nontrivial fusion rules, can be directly probed through the power-law scaling exponent of nonlinear signal near the threshold frequency for anyon-pair creation.
We analytically showed that along one direction in the frequency plane, the exchange phases between an anyon $a$ and its antiparticle $\overline{a}$ can be measured, while along a second direction, the fractional exchange phase between two anyons $b$ and $c$ forming the composite topological charge $\overline{a}$ can be extracted. These findings extend the previous theoretical work demonstrating that linear threshold spectroscopy can access the fractional exchange statistics of an anyon-antianyon pair $a\overline{a}$~\cite{PhysRevLett.118.227201}.

The analytic predictions were numerically verified using exact diagonalization to simulate the nonlinear dynamics of anyonic Hamiltonians~\cite{PhysRevB.43.2661, PhysRevB.43.10761, PhysRevB.107.195129}. For the toric code anyons~\cite{KITAEV20032, 1506.05805, topologicalquantum}, the mutual semionic exchange statistics between the electric and magnetic charges were found to agree with the theoretical predictions, while for Ising anyons~\cite{KITAEV20062, topologicalquantum}, the non-Abelian braiding statistics in the fermionic fusion channel was confirmed.

Because our results assume that the physical system, including all relevant low-energy excitations, is described by an effective anyon model without reference to any specific microscopic realization of topological order, a natural follow-up question is how our theoretical predictions can be verified in concrete and realistic systems such as quantum spin liquids or topological superconductors. The effects of disorder and finite temperature on the spectral line shape would also be important to assess the practical applicability of our proposed protocol. Such studies would then motivate experimental implementations to test these theoretical predictions.

\section*{Acknowledgements}

The authors thank Adam Gammon-Smith for helpful discussions.
N.K. and F.P. acknowledge support from the Deutsche Forschungsgemeinschaft
(DFG, German Research Foundation) under Germany's Excellence Strategy EX-2111-390814868, TRR 360 - 492547816, FOR 5522 (project-id 499180199), the European Research Council (ERC) under the European Union's Horizon 2020 research and innovation programme (Grant Agreement No. 771537), as well as the Munich Quantum Valley, which is supported by the Bavarian state government with funds from the Hightech Agenda Bayern Plus.
W.C. is supported at Los Alamos National Laboratory (LANL) which is operated by Triad National Security, LLC, for the National Nuclear Security Administration of U.S. Department of Energy (Contract No. 89233218CNA000001).
W.C. also gratefully acknowledges the support of the U.S. Department of Energy through the LANL/LDRD Program and the Center for Nonlinear Studies for this work.

\section*{Data Availability}

Access to the simulation codes and numerical data shown in the figures may be granted upon reasonable request~\cite{zenodo}.

\appendix

\begin{widetext}

\section{Derivation of the threshold behavior of $\mathcal{C}$}
\label{app:chi3}

The aim of this section is to derive the result for the threshold behavior of $\mathcal{C}(T,\omega_{\tau},\omega_{t})$ in Eq.~\eqref{eq:chi3_threshold} by utilizing the assumptions discussed in the main text in Sec.~\ref{sec:anyon2DCS}. The process we focus on is where the anyon-antianyon pair $a\overline{a}$ is first excited on top of the ground state and propagates for time $T$, followed by splitting of $\overline{a}$ into $b$ and $c$, which again recombine to $\overline{a}$ after time $\tau$. The anyons $a$ and $\overline{a}$ finally annihilate again to the vacuum after time $t$,
\begin{align}
	|0\rangle \langle 0| \longrightarrow  |{a\overline{a}} \rangle \langle 0| \xrightarrow{\,\,  T \,\,}  |{abc}\rangle \langle 0| \xrightarrow{\,\, \tau \,\,\,}  |{a\overline{a}} \rangle \langle 0| \xrightarrow{\,\, t \,\,\,}  |0\rangle\langle 0|  
\end{align}
(cf. Eq.~\eqref{eq:pathway}). The wave functions $|{a\overline{a}} \rangle$ and $|{abc}\rangle$ are given by Eqs.~\eqref{eq:2anyonWF} and \eqref{eq:3anyonWF} in the main text. In a final step, $\mathcal{C}(T,\tau,t)$ is Fourier transformed with respect to $\tau$ and $t$ to obatin $\mathcal{C}(T,\omega_{\tau},\omega_{t})$. Note that the above process is the only relevant contribution to the susceptibility $\chi^{(3)}(T,\omega_{\tau},\omega_t)$ in the limit and region in the $\omega_{\tau}-\omega_t$ plane we are interested in. In the following, we first consider some side computations of matrix elements that are used later on.

\paragraph*{$|0\rangle \rightarrow  |{a\overline{a}} \rangle$:} The matrix element $\mathcal{M}_1^{a\overline{a}}$ connecting the ground state $|0\rangle$ and the anyon pair state $|{a\overline{a}} \rangle$ can be evaluated as done in Ref.~\cite{PhysRevLett.118.227201} by making use of the two-anyon wave function Eq.~\eqref{eq:2anyonWF} and the action of the drive pulses Eq.~\eqref{eq:drive}:
\begin{align}
	\mathcal{M}_1^{a\overline{a}}(l, \mathbf{K}, k) &\propto \int \mathrm{d}\mathbf{R}\int \mathrm{d}\phi \sum_m c_m e^{im\phi}  \langle {a\overline{a}}| \mathcal{O}(\mathbf{R},\dist,\phi) |0\rangle\\
	&\propto M_1^{a\overline{a}}  \sum_m \int\mathrm{d}\phi \, c_m e^{i(m+\alpha-l)\phi} \sqrt{\frac{k}{L^3}} J_{|l-\alpha|}(k\dist) \int \mathrm{d}\mathbf{R}\, e^{-i\mathbf{K}\cdot \mathbf{R}}\\
	&\propto M_1^{a\overline{a}} c(\alpha - l) \sqrt{kL} J_{|l-\alpha|}(k\dist)  \delta_{\mathbf{K},0},
\label{eq:matrixelement2}
\end{align}
where we have introduced
\begin{align}
	c(\alpha) = \sum_m \int\mathrm{d}\phi \, c_m e^{i(m+\alpha)\phi} = \sum_m c_m \frac{e^{2\pi i \alpha} - 1}{i(m+\alpha)} \quad \text{for $\alpha\notin \mathbb{Z}$}
\end{align}
for convenience. The case $\alpha\in \mathbb{Z}$, for which $c(\alpha)=2\pi c_{-\alpha}$, is only relevant when considering anyons with trivial exchange statistics. Note that we are only interested in the frequency dependence of the final result $\mathcal{C}(T,\omega_{\tau},\omega_{t})$, such that constants can be dropped, as already done in Eq.~\eqref{eq:2anyonWF}. The $\delta_{\mathbf{K},0}$ in Eq.~\eqref{eq:matrixelement2} reflects the assumption that there is no total momentum transfer between the drive and the system. When considering the above matrix element for the computation of $\mathcal{C}$, all possible values of the angular momentum $l$ need to be considered.

\paragraph*{$|{a\overline{a}} \rangle \rightarrow | {abc} \rangle$:} The matrix element $\mathcal{M}_{\overline{a}}^{bc}$ connecting the anyon pair state $|{a\overline{a}} \rangle$ and the anyon triplet state $|{abc}\rangle$ is evaluated by exploiting the assumption that the three-anyon wave function can be approximated as product state of a plane wave and a two-anyon wave function (Eq.~\eqref{eq:3anyonWF}):
\begin{align}
	\mathcal{M}_{\overline{a}}^{bc}(l', \mathbf{k}_a, \mathbf{K}', k' ; l, \mathbf{K}, k) \propto&  \int \mathrm{d}\mathbf{R}'\int \mathrm{d}\phi' \sum_m c_m e^{im\phi'}  \langle {abc}| \mathcal{O}(\mathbf{R}',\dist,\phi') | {a\overline{a}} \rangle  \label{eq:matrixelement3-0}\\
\begin{split}
	\propto& M_{\overline{a}}^{bc} \int \mathrm{d}\mathbf{R}'\int \mathrm{d}\phi' \sum_m c_m e^{im\phi'} \frac{\sqrt{kk'}}{L^4} \int\mathrm{d}\mathbf{R} \int\mathrm{d}r\,r \int \mathrm{d}\phi \,  J_{|l'-\beta|}(k'\dist)e^{-i(l'-\beta)\phi'}\\
	&\quad\times e^{-i\mathbf{K}'\cdot \mathbf{R}'}e^{-i\mathbf{k}_a\cdot \mathbf{r}_a} J_{|l-\alpha|}(kr)e^{i(l-\alpha)\phi}e^{i\mathbf{K}\cdot \mathbf{R}} \delta(\mathbf{R}'-\mathbf{r}_{\overline{a}}).
\end{split}
\label{eq:matrixelement3-1}
\end{align}
In this expression, the primed quantities are associated with the two-anyon wave function $|{bc} \rangle$. In particular, $l'$ is the relative angular momentum, $\phi'$ the relative orientation, $k'$ the relative momentum, and $\mathbf{K}'$ and $\mathbf{R}'$ the center of mass momentum and position, respectively. The unprimed symbols describe the corresponding quantities for the two-anyon wave function $|{a\overline{a}} \rangle$, just like above for the matrix element $\mathcal{M}_1^{a\overline{a}}$. Further, $r=|\mathbf{r}_a - \mathbf{r}_{\overline{a}}|$ is the distance between $a$ and $\overline{a}$, whose positions are $\mathbf{r}_a$ and $\mathbf{r}_{\overline{a}}$, respectively; $\mathbf{k}_a$ denotes the momentum of anyon $a$ in the three-anyon wave function $| {abc} \rangle$. The statistical parameters $\alpha$ and $\beta$ describe the anyonic exchange phases associated with $R^{a\overline{a}}_1$ and $R^{bc}_{\overline{a}}$, respectively. The $\delta$-function in Eq.~\eqref{eq:matrixelement3-1} arises due to the pulse splitting $\overline{a}$ into anyons $b$ and $c$, implying that the operator $\mathcal{O}$ must act at location $\mathbf{r}_{\overline{a}}$. The $\mathbf{R}$ integral can be evaluated by first performing the trivial $\mathbf{R}'$ integration and using
\begin{align}
	\mathbf{r}_{\overline{a}}= \mathbf{R} + \frac{m_ar}{m_{{a}} + m_{\overline{a}}} \mathbf{e}_r, \quad
	\mathbf{r}_{a}= \mathbf{R} - \frac{m_{\overline{a}} r}{m_{{a}} + m_{\overline{a}}} \mathbf{e}_r, \quad \text{with $\mathbf{e}_r = \begin{pmatrix} \cos(\phi) \\ \sin(\phi)\end{pmatrix}$},
\end{align}
where it was utilized that $\mathbf{R}=(m_a\mathbf{r}_{{a}} + m_{\overline{a}}\mathbf{r}_{\overline{a}})/(m_{{a}} + m_{\overline{a}})$ and $\mathbf{r}_{\overline{a}} - \mathbf{r}_a = r\mathbf{e}_r$. We then have
\begin{align}
\begin{split}
	\mathcal{M}_{\overline{a}}^{bc} (l', \mathbf{k}_a, \mathbf{K}', k' &; l, \mathbf{K}, k)
	\propto M_{\overline{a}}^{bc} \int \mathrm{d} \phi' \sum_m c_m e^{i(m +\beta -l')\phi'} \frac{\sqrt{kk'}}{L^4} J_{|l'-\beta|}(k'\dist) \int\mathrm{d}\mathbf{R} \int\mathrm{d}r\,r \int \mathrm{d}\phi \,     \\
	&\quad\quad\quad\quad\quad \times e^{-i\mathbf{K}'\cdot (  \mathbf{R} + \frac{m_a r}{m_{{a}} + m_{\overline{a}}} \mathbf{e}_r  )}  e^{-i\mathbf{k}_a\cdot ( \mathbf{R} - \frac{m_{\overline{a}}r}{m_{{a}} + m_{\overline{a}}} \mathbf{e}_r )} J_{|l-\alpha|}(kr)e^{i(l-\alpha)\phi}e^{i\mathbf{K}\cdot \mathbf{R}}
\end{split}\\
\begin{split}
	\propto& M_{\overline{a}}^{bc} c(\beta -l') \frac{\sqrt{kk'}}{L^2} J_{|l'-\beta|}(k'\dist) \int\mathrm{d}r\,r \int \mathrm{d}\phi \,     e^{-i\frac{m_a r}{m_{{a}} + m_{\overline{a}}}\mathbf{K}\cdot \mathbf{e}_r} e^{ir\mathbf{k}_a \cdot \mathbf{e}_r }   J_{|l-\alpha|}(kr)e^{i(l-\alpha)\phi}  \delta_{\mathbf{K},\mathbf{K}'+\mathbf{k}_{{a}}}.
\end{split}
\label{eq:matrixelement3-2}
\end{align}
Going forward, we are free to assume that $\mathbf{K}=0$ since $\mathbf{K}$ is the total momentum that is assumed to be conserved and trivial (cf. Eq.~\eqref{eq:matrixelement2}). The scalar product $\mathbf{k}_a \cdot \mathbf{e}_r$ can be reexpressed as
\begin{align}
	\mathbf{k}_a \cdot \mathbf{e}_r = k_{a,x}\cos(\phi) + k_{a,y}\sin(\phi) = k_a \sin(\phi-\phi_a),
\end{align}
where $\mathbf{k}_a = (k_{a,x}, k_{a,y})^{\top}$, $|\mathbf{k}_a| = k_a$, $\cos(\phi_a)=k_{a,y} / k_a$, and $\sin(\phi_a)=-k_{a,x} / k_a$. Using this relation, the $\phi$ integration in Eq.~\eqref{eq:matrixelement3-2} can be evaluated to be
\begin{align}
	\int_0^{2\pi} \mathrm{d}\phi \,  e^{ik_ar\sin(\phi - \phi_a) + i(l-\alpha)\phi} &= e^{i(l-\alpha)\phi_a}\int_{-\phi_a}^{2\pi-\phi_a} \mathrm{d}\phi \,  e^{ik_ar\sin(\phi) + i(l-\alpha)\phi}\\
	&= e^{i(l-\alpha)\phi_a} \int_{-\phi_a}^{2\pi-\phi_a} \mathrm{d}\phi \sum_{n=-\infty}^{\infty} J_n(k_a r) e^{i n\phi} e^{i(l-\alpha)\phi}\\
	&=
	\begin{cases}
	\displaystyle 2\sin(\alpha\pi) e^{-i\pi\alpha}\sum_{n=-\infty}^{\infty}\frac{e^{-in\phi_a}}{\alpha-n-l}J_n(k_ar) \quad &\text{if $\alpha \notin \mathbb{Z}$,} \\
	2\pi e^{i(l-\alpha)\phi_a} J_{\alpha - l}(k_a r) \quad &\text{if $\alpha \in \mathbb{Z}$,}
	\end{cases}
	\label{eq:matrixelement3-phiint}
\end{align}
where it was used that $ik_ar\sin(\phi) = k_ar (e^{i\phi} - e^{-i\phi}) /2$ and that the generating function for the Bessel functions of the first kind is
\begin{align}
	e^{\frac{x}{2}(z-z^{-1})} = \sum_{n=-\infty}^{\infty} J_n(x)z^n.
\end{align}
We will focus on the more general case $\alpha \notin \mathbb{Z}$ in the following; adapting the result to the other case is straight forward. Plugging Eq.~\eqref{eq:matrixelement3-phiint} into Eq.~\eqref{eq:matrixelement3-2} yields the expression
\begin{align}
	\mathcal{M}_{\overline{a}}^{bc}
	\propto& M_{\overline{a}}^{bc} c(\beta -l') \frac{\sqrt{kk'}}{L^2} J_{|l'-\beta|}(k'\dist) 2\sin(\alpha\pi) e^{-i\pi\alpha}\sum_n\frac{e^{-in\phi_a}}{\alpha-n-l} \int\mathrm{d}r\,r  J_n(k_ar)   J_{|l-\alpha|}(kr)  \delta_{\mathbf{K},\mathbf{K}'+\mathbf{k}_{{a}}},
	\label{eq:matrixelement3-3}
\end{align}
which features a so-called Weber-Schafheitlin-type integral with the solution~\cite{Kellendonk01022009, 1004.5518}
\begin{align}
\begin{split}
	\mathcal{I}_{\mu,\nu}(s) \equiv \int_0^{\infty} \mathrm{d} x \, xJ_\mu(sx)J_{\nu}&(x) = \cos\left(\pi\frac{\mu - \nu}{2}\right)\delta(s-1) + \frac{2}{\pi}\sin\left(\pi\frac{\nu-\mu}{2}\right)  \mathcal{P}\left(\frac{1}{1/s - s}\right)\\
	&\times \begin{cases}
                    s^{\mu -1} \frac{\Gamma(\frac{\mu + \nu}{2}+1)\Gamma(\frac{\mu - \nu}{2}+1)}{\Gamma(\mu+1)}  {}_2F_1(\frac{\mu + \nu}{2},\frac{\mu - \nu}{2};\mu + 1; s^2) & \text{if $s\leq 1$, $\mu + 2 > |\nu|$,}  \\
                    s^{-\nu -1} \frac{\Gamma(\frac{\mu + \nu}{2}+1)\Gamma(\frac{\nu - \mu}{2}+1)}{\Gamma(\nu+1)} {}_2F_1(\frac{\mu + \nu}{2},\frac{\nu - \mu}{2};\nu + 1; s^{-2}) & \text{if $s> 1$, $\nu + 2 > |\mu|$,}
                 \end{cases}
\end{split}
\label{eq:WS-integral}
\end{align}
where $\mathcal{P}$ denotes the principal value, ${}_2F_1(a,b;c;z)$ the hypergeometric function, and it is assumed that $s \in \mathbb{R}_{+}$. The above result can be extended beyond the specified conditions in $s$ by analytic continuation~\cite{BECKEN2000449}, such that we have a solution for the integral $\mathcal{I}_{\mu,\nu}(s)$ for all $s\in \mathbb{R}_+$ and both $\mu + 2 > |\nu|$ and $\nu + 2 > |\mu|$. In order to fit the solution in Eq.~\eqref{eq:WS-integral} into our normalization scheme, one may transform the $\delta$-function as
\begin{align}
	\delta(s-1)=\delta\left(\frac{k_a}{k}-1\right) = k\delta(k_a-k) \longrightarrow \frac{kL}{2\pi}\delta_{k_a,k},
\end{align}
where it was used that the dimensionless parameter is $s = k_a/k$. In practice, we do not need to apply the above transform to Eq.~\eqref{eq:WS-integral} since in the final step of computing $\mathcal{C}$, the summations over the different momenta can be transformed to integrals by taking the continuum limit. Overall, the matrix element Eq.~\eqref{eq:matrixelement3-0} is given by
\begin{align}
	\mathcal{M}_{\overline{a}}^{bc}
	\propto& 2\sin(\alpha\pi) e^{-i\pi\alpha} M_{\overline{a}}^{bc} c(\beta -l') \frac{\sqrt{kk'}}{k^2L^2} J_{|l'-\beta|}(k'\dist)  \sum_n\frac{e^{-in\phi_a}}{\alpha-n-l} \mathcal{I}_{n,|l-\alpha|}\left( \frac{k_a}{k} \right)  \delta_{\mathbf{K},\mathbf{K}'+\mathbf{k}_{{a}}}.
	\label{eq:matrixelement3-4}
\end{align}

\paragraph*{$\mathcal{C}(T,\omega_{\tau},\omega_{t})$:} Using the results for the matrix elements in Eqs.~\eqref{eq:matrixelement2} and \eqref{eq:matrixelement3-4}, we can now compute the contribution $\mathcal{C}(T,\omega_{\tau},\omega_{t})$ to $\chi^{(3)}(T,\tau,t)$ we are interested in and then finally calculate its Fourier transform $\mathcal{C}(T,\omega_{\tau},\omega_{t})$:
\begin{align}
\begin{split}
	\mathcal{C}(T,\tau,t) =  \frac{i^3}{L^2}\Theta(t)\Theta(\tau)\Theta(T) &\sum_{l,l',l''}\sum_{\mathbf{K}, \mathbf{K'}, \mathbf{K''}}\sum_{\mathbf{k}_a}\sum_{k,k',k''}     e^{-i\widetilde{E}_{a\overline{a}}t} e^{-iE_{abc}\tau}  e^{-iE_{a\overline{a}}T}       \mathcal{M}_1^{a\overline{a}}(l'', \mathbf{K}'', k'')^* \\  
	& \times \mathcal{M}_{\overline{a}}^{bc} (l', \mathbf{k}_a, \mathbf{K}', k' ; l'', \mathbf{K}'', k'')^* \mathcal{M}_{\overline{a}}^{bc} (l', \mathbf{k}_a, \mathbf{K}', k' ; l, \mathbf{K}, k)          \mathcal{M}_1^{a\overline{a}}(l, \mathbf{K}, k),
\end{split}
\label{eq:chi3_def}
\end{align}
where $E_{a\overline{a}}=k^2/(2\mu_{a\overline{a}}) + \Delta_{a\overline{a}}$ is the energy of the two-anyon state $|{a\overline{a}} \rangle$ that is excited at time $0$, $E_{abc}=k_a^2/(2m_a) + k_a^2/(2(m_b+m_c)) + k'^2/(2\mu_{bc}) + \Delta_{abc}$ the energy of the three-anyon state $|{abc}\rangle$ that is excited at time $T$, and $\widetilde{E}_{a\overline{a}}=k''^2/(2\mu_{a\overline{a}}) + \Delta_{a\overline{a}}$ the energy of the two-anyon state $|{a\overline{a}} \rangle$ that is excited at time $T+\tau$, with $\Delta$ denoting the spectral gaps for creating the anyons indicated in the respective indices. Note that we suppress the sum over all different anyon species $a,b,c$ in Eq.~\eqref{eq:chi3_def} for convenience and continue doing so in the remainder of this section. Since the product of the matrix elements in Eq.~\eqref{eq:chi3_def} is proportional to $\delta_{\mathbf{K''},0} \delta_{\mathbf{K}'',\mathbf{K}'+\mathbf{k}_a}  \delta_{\mathbf{K},\mathbf{K}'+\mathbf{k}_a}   \delta_{\mathbf{K},0}$ (cf. Eqs.~\eqref{eq:matrixelement2} and \eqref{eq:matrixelement3-4}), the sums over $\mathbf{K}, \mathbf{K'}, \mathbf{K''}$ become trivial and we set $\mathbf{K}'=-\mathbf{k}_a$, such that
\begin{align}
\begin{split}
	\mathcal{C}(T,\tau,t) \propto \frac{i^3}{L^2}\Theta(t)\Theta(\tau)\Theta(T) &\sum_{l,l',l''}\sum_{\mathbf{k}_a}\sum_{k,k',k''}    4 \sin^2(\alpha \pi)  |M_1^{a\overline{a}}|^2  |M_{\overline{a}}^{bc}|^2  c(\alpha -l'')^*  |c(\beta -l')|^2  c(\alpha -l)\\
	  &\times e^{-i\widetilde{E}_{a\overline{a}}t} e^{-iE_{abc}\tau}  e^{-iE_{a\overline{a}}T} 
	  \frac{k'}{kk''L^3} J_{|l''-\alpha|}(k''\dist) J_{|l'-\beta|}^2(k'\dist) J_{|l-\alpha|}(k\dist)\\
	  &\times \sum_{n,m} \frac{e^{-i(n-m)\phi_a}}{(\alpha-n-l)(\alpha-m-l'')} \mathcal{I}_{n,|l-\alpha|}\left( \frac{k_a}{k} \right) \mathcal{I}_{m,|l''-\alpha|}\left( \frac{k_a}{k''} \right).
\end{split}
\end{align}
Taking the continuum limit for the sums over $k$, $k'$, $k''$, and $\mathbf{k}_a$, with $\sum_{\mathbf{k}_a} \longrightarrow (L/(2\pi))^2\int \mathrm{d}k_a\,k_a \int \mathrm{d}\phi_a$, and carrying out the $\phi_a$ integration yields
\begin{align}
\begin{split}
	\mathcal{C}(T,\tau,t) \propto \frac{i^3}{8\pi^5} \sin^2(\alpha \pi) & |M_1^{a\overline{a}}|^2  |M_{\overline{a}}^{bc}|^2 \Theta(t)\Theta(\tau)\Theta(T) \sum_{l,l',l''}   c(\alpha -l'')^*  |c(\beta -l')|^2  c(\alpha -l)\\
	  & \times \int \mathrm{d}k \int \mathrm{d}k' \int \mathrm{d}k'' \int \mathrm{d}k_a\, e^{-i\widetilde{E}_{a\overline{a}}t} e^{-iE_{abc}\tau}  e^{-iE_{a\overline{a}}T}
	  \frac{k'k_a}{kk''} J_{|l''-\alpha|}(k''\dist)J_{|l'-\beta|}^2(k'\dist) \\
	  &\times  J_{|l-\alpha|}(k\dist) \sum_{n} \frac{1}{(\alpha-n-l)(\alpha-n-l'')} \mathcal{I}_{n,|l-\alpha|}\left( \frac{k_a}{k} \right) \mathcal{I}_{n,|l''-\alpha|}\left( \frac{k_a}{k''} \right).
\end{split}
\end{align}
Thus, the Fourier transform $\mathcal{C}(T,\omega_{\tau},\omega_{t})$ can be expressed as
\begin{align}
\begin{split}
	\mathcal{C}(T,\omega_{\tau},\omega_{t}) &=  \int \mathrm{d}t \int \mathrm{d}\tau\, e^{i\omega_{\tau} \tau} e^{i\omega_{t} t}  \mathcal{C}(T,\tau,t)\\
	&\propto \frac{i^3}{8\pi^5} \sin^2(\alpha \pi) |M_1^{a\overline{a}}|^2  |M_{\overline{a}}^{bc}|^2 \Theta(\Omega_{t})\Theta(\Omega_{\tau})\Theta(T) \sum_{l,l',l''}   c(\alpha -l'')^*  |c(\beta -l')|^2  c(\alpha -l)\\
	& \times \int \mathrm{d}k \int \mathrm{d}k' \int \mathrm{d}k'' \int \mathrm{d}k_a\, \left( \pi \delta \left(\omega_{t} - \widetilde{E}_{a\overline{a}}\right)  + \frac{i}{\omega_{t} - \widetilde{E}_{a\overline{a}}} \right)  \left( \pi \delta \left(\omega_{\tau} - E_{abc}\right)  + \frac{i}{\omega_{\tau} - E_{abc}} \right)  e^{-iE_{a\overline{a}}T}\\
	&\times  \frac{k'k_a}{kk''} J_{|l''-\alpha|}(k''\dist)J_{|l'-\beta|}^2(k'\dist) J_{|l-\alpha|}(k\dist) \sum_{n} \frac{1}{(\alpha-n-l)(\alpha-n-l'')} \mathcal{I}_{n,|l-\alpha|}\left( \frac{k_a}{k} \right) \mathcal{I}_{n,|l''-\alpha|}\left( \frac{k_a}{k''} \right),
\end{split}
\end{align}
where it was used that infinitesimal imaginary parts $\epsilon_{\tau}$ and $\epsilon_{t}$ can be added to $\omega_{\tau}$ and $\omega_{t}$, that is, $\omega_{\tau} \longrightarrow \omega_{\tau} + i\epsilon_{\tau}$ and $\omega_{t} \longrightarrow \omega_{t} + i\epsilon_{t}$, for which the limit $\epsilon_{\tau}, \epsilon_{t}\longrightarrow 0^+$ is taken. We further introduced the frequencies above the spectral gaps $\Omega_{\tau} = \omega_{\tau} - \Delta_{abc}$ and $\Omega_{t} = \omega_{t} - \Delta_{a\overline{a}}$. It is generally expected that the contributions arising from the terms $(\omega_{t} - \widetilde{E}_{a\overline{a}})^{-1}$ and $(\omega_{\tau} - E_{abc})^{-1}$ lead to broadening effects in frequency space that do not affect the observed threshold behavior of $\mathcal{C}$. Neglecting these contributions, the integrations over $k'$ and $k''$ become trivial due to the $\delta$-functions, which select the momenta $k'=\sqrt{2\mu_{bc}\Omega_{\tau} - \mu_{bc}k_a^2(m_a+m_b+m_c)/(m_a(m_b+m_c))}$ and $k''=\sqrt{2\mu_{a\overline{a}}\Omega_{t}}$,
\begin{align}
\begin{split}
	\mathcal{C}(T,\omega_{\tau},\omega_{t}) &\propto \frac{i^3}{8\pi^3} \sin^2(\alpha \pi) |M_1^{a\overline{a}}|^2  |M_{\overline{a}}^{bc}|^2 \Theta(\Omega_{t})\Theta(\Omega_{\tau})\Theta(T) \sum_{l,l',l''}   c(\alpha -l'')^*  |c(\beta -l')|^2  c(\alpha -l)\\
	& \times \int \mathrm{d}k \int \mathrm{d}k_a\, \mu_{a\overline{a}}\mu_{bc} e^{-iE_{a\overline{a}}T}  \frac{k_a}{k\sqrt{2\mu_{a\overline{a}}\Omega_{t}}^2} J_{|l''-\alpha|}\left( \sqrt{2\mu_{a\overline{a}}\Omega_{t}} \dist \right) J_{|l'-\beta|}^2\left(\sqrt{2\mu_{bc}\Omega_{\tau} \dist^2 - \frac{\mu_{bc}}{\mu_{a\overline{a}}}k_a^2\dist \dist'}\right) \\
	&\times  J_{|l-\alpha|}(k\dist) \sum_{n} \frac{1}{(\alpha-n-l)(\alpha-n-l'')} \mathcal{I}_{n,|l-\alpha|}\left( \frac{k_a}{k} \right) \mathcal{I}_{n,|l''-\alpha|}\left( \frac{k_a}{\sqrt{2\mu_{a\overline{a}}\Omega_{t}}} \right)
	\Theta(k_{a,\text{max}} - k_a),
\end{split}
\label{eq:chi3_calc}
\end{align}
where we introduced the rescaled distance 
\begin{align}
	\dist' &= \dist \frac{\mu_{a\overline{a}}}{m_a} \frac{m_a+m_b+m_c}{m_b+m_c}.
\end{align}
Since the momentum $k'$ is non-negative, there is an upper bound for the $k_a$ integration given by
\begin{align}
	k_{a,\text{max}} = \sqrt{2\Omega_{\tau}\frac{m_a(m_b+m_c)}{m_a+m_b+m_c}}
	\label{eq:ka_upperbound}
\end{align}
The non-negativity condition in $k'$ can thus be expressed as the additional factor
\begin{align}
	\Theta\left(\Omega_{\tau} - \frac{\dist'}{\dist}\frac{k_a^2}{2\mu_{a\overline{a}}} \right) = \Theta(k_{a,\text{max}} - k_a).
\end{align}
that can be found in Eq.~\eqref{eq:chi3_calc}. In the next step, we argue that in Eq.~\eqref{eq:chi3_calc}, only the contribution proportional to the $\delta$-functions in $\mathcal{I}_{\mu,\nu}$ (cf. Eq.~\eqref{eq:WS-integral}) are relevant.

Since our main interest is the threshold behavior, we consider the limit of small $\Omega_{\tau}$ and $\Omega_{t}$. In the limit $\Omega_{\tau}\longrightarrow 0$, the upper bound for the $k_a$ integral in Eq.~\eqref{eq:ka_upperbound} approaches zero, $k_{a,\text{max}} \longrightarrow 0$, such that the integral reduces to a single point. Since the second term in $\mathcal{I}_{\mu,\nu}(s)$ is well behaved for every $s\in \mathbb{R}_+$~\cite{Kellendonk01022009}, its contribution to the integral must be vanishing in this limit, which implies that only the term proportional to the $\delta$-function is of importance.

One of the central assumptions in Sec.~\ref{sec:anyon2DCS} and in the above calculation is that the three-anyon wave function $|{abc}\rangle$ can be approximated as product state of a plane wave in $a$ and a two-anyon wave function in $b$ and $c$, which is expected to be valid when the splitting $\overline{a}\rightarrow bc$ happens while $a$ and $\overline{a}$ are spatially separated. To ensure the validity of this condition for small $k$ (the relative momentum between $a$ and $\overline{a}$), the initial time step $T$ must be chosen to be large. This implies that the term $\exp(-iE_{a\overline{a}}T)$ with $E_{a\overline{a}}=k^2/(2\mu_{a\overline{a}}) + \Delta_{a\overline{a}}$ strongly oscillates when varying $k$. Since the non-$\delta$-contribution to $\mathcal{I}_{n,|l-\alpha|}( k_a/k)$ is a smooth function (as is the remaining functions in Eq.~\eqref{eq:chi3_calc}), the strong oscillations are expected to lead to a result that is negligible compared to the contribution arising from the $\delta$-function in $\mathcal{I}_{n,|l-\alpha|}( k_a/k)$.

We can therefore neglect the contributions in $\mathcal{I}_{n,|l-\alpha|}( k_a/k ) \mathcal{I}_{n,|l''-\alpha|}( k_a/\sqrt{2\mu_{a\overline{a}}\Omega_{t}} )$ that are not proportional to $\delta$-functions, such that we obtain $k=k_a=\sqrt{2\mu_{a\overline{a}}\Omega_{t}}$ and
\begin{align}
\begin{split}
	\mathcal{C}(T,\omega_{\tau},\omega_{t}) &\propto \frac{i^3}{8\pi^3} \sin^2(\alpha \pi) |M_1^{a\overline{a}}|^2  |M_{\overline{a}}^{bc}|^2 \Theta(\Omega_{t})\Theta(\Omega_{\tau})\Theta(T)\Theta\left( \Omega_{\tau} - \frac{\dist'}{\dist}\Omega_{t} \right) \sum_{l,l',l''}   c(\alpha -l'')^*  |c(\beta -l')|^2  c(\alpha -l)\\
	& \times \mu_{a\overline{a}}\mu_{bc} e^{-iE_{a\overline{a}}T}  J_{|l''-\alpha|}\left( \sqrt{2\mu_{a\overline{a}}\Omega_{t}} \dist \right) J_{|l'-\beta|}^2\left(\sqrt{2\mu_{bc}\Omega_{\tau}\dist^2 - 2\mu_{bc}\Omega_{t}\dist\dist'}\right) J_{|l-\alpha|}\left( \sqrt{2\mu_{a\overline{a}}\Omega_{t}}\dist \right) \\
	&\times  \sum_{n} \frac{1}{(\alpha-n-l)(\alpha-n-l'')} \cos\left(\pi \frac{|n|-|l-\alpha|}{2}  \right)  \cos\left(\pi \frac{|n|-|l''-\alpha|}{2}  \right).
\end{split}
\end{align}
As we focus on the threshold behavior, the Bessel functions can now be expanded up to leading order in $\Omega_{\tau}$ and $\Omega_{t}$. Note that the statistical parameters $\alpha$ and $\beta$, which encode the exchange phases between $a$ and $\overline{a}$, and between $b$ and $c$, can always be chosen to be $\alpha, \beta \in (-1,1]$, such that the most relevant contributions come from the case $l=l'=l''=0$. We thus find
\begin{align}
	|\mathcal{C}(T,\omega_{\tau},\omega_{t})| &\propto |M_1^{a\overline{a}}|^2  |M_{\overline{a}}^{bc}|^2   |c(\alpha)|^2  |c(\beta )|^2   (\dist^2 \Omega_{t})^{|\alpha|} (\dist^2 \Omega_{\tau} - \dist\dist' \Omega_{t})^{|\beta|}      \Theta(\Omega_{t})\Theta(T)\Theta\left( \dist\Omega_{\tau} - \dist'\Omega_{t} \right),
	\label{eq:chi3_threshold_app}
\end{align}
where it was used that the condition $\Theta(\Omega_{\tau})$ is already enforced by $\Theta(\Omega_{t})\Theta ( \dist\Omega_{\tau} - \dist'\Omega_{t} )$, such that the factor $\Theta(\Omega_{t})$ can be dropped.
Note that there is an alternate, semiclassical argument why only the $\delta$-functions in the two $\mathcal{I}_{\mu,\nu}$ terms are relevant for the leading order of $\mathcal{C}$. When the splitting process $\overline{a}\rightarrow bc$ occurs while $a$ and $\overline{a}$ are sufficiently separated, the momentum of $a$ semiclassically remains unaffected due to the local nature of the splitting process. In terms of the convention used for $\mathcal{C}$ (cf. Eq.~\eqref{eq:chi3_def}), this implies that $k=k_a=k''$, which is precisely the case that is selected by the $\delta$-functions in $\mathcal{I}_{n,|l-\alpha|}( k_a/k ) \mathcal{I}_{n,|l''-\alpha|}( k_a/k'')$. The contributions coming from $k\neq k_a\neq k''$ are thus expected to be subleading corrections in the limit we are interested in.

The above intuition can also be used to find appropriate protocols for actual experiments (as $T \longrightarrow \infty$ and $\Omega_{\tau},\Omega_{t}\longrightarrow 0$ are experimentally unachievable). The derived threshold behavior Eq.~\eqref{eq:chi3_threshold_app} is expected to hold when the energy of the initial pulse and the time $T$ are chosen such that $a$ and $\overline{a}$ are well-separated when the second pulse reaches the system.

\end{widetext}

\bibliography{./references.bib}

@dataset{zenodo,
author = {Nico Kirchner and Wonjune Choi and Frank Pollmann},
title = "{Measuring Anyonic Exchange Phases Using Two-Dimensional Coherent Spectroscopy}",
month = {11},
year = {2025},
publisher = {Zenodo},
doi = {10.5281/zenodo.17659262},
url = {https://doi.org/10.5281/zenodo.17659262},
note = "{Zenodo}",
}

@article{PhysRevLett.118.227201,
  title = "{Statistics of Fractionalized Excitations through Threshold Spectroscopy}",
  author = {Morampudi, Siddhardh C. and Turner, Ari M. and Pollmann, Frank and Wilczek, Frank},
  journal = {Phys. Rev. Lett.},
  volume = {118},
  issue = {22},
  pages = {227201},
  numpages = {5},
  year = {2017},
  month = {May},
  publisher = {American Physical Society},
  doi = {10.1103/PhysRevLett.118.227201},
  url = {https://link.aps.org/doi/10.1103/PhysRevLett.118.227201}
}

@article{BONDERSON20194065,
title = {On invariants of modular categories beyond modular data},
journal = {Journal of Pure and Applied Algebra},
volume = {223},
number = {9},
pages = {4065-4088},
year = {2019},
issn = {0022-4049},
doi = {https://doi.org/10.1016/j.jpaa.2018.12.017},
url = {https://www.sciencedirect.com/science/article/pii/S0022404918303098},
author = {Parsa Bonderson and Colleen Delaney and César Galindo and Eric C. Rowell and Alan Tran and Zhenghan Wang},
}

@article{PhysRevResearch.3.033110,
  title = {Measuring topological order},
  author = {Bonderson, Parsa},
  journal = {Phys. Rev. Res.},
  volume = {3},
  issue = {3},
  pages = {033110},
  numpages = {29},
  year = {2021},
  month = {Aug},
  publisher = {American Physical Society},
  doi = {10.1103/PhysRevResearch.3.033110},
  url = {https://link.aps.org/doi/10.1103/PhysRevResearch.3.033110}
}

@book{topologicalquantum,
    author = {Simon, Steven H.},
    title = "{Topological Quantum}",
    publisher = {Oxford University Press},
    year = {2023},
    month = {09},
    isbn = {9780198886723},
    doi = {10.1093/oso/9780198886723.001.0001},
    url = {https://doi.org/10.1093/oso/9780198886723.001.0001},
}

@article{1506.05805,
  title={Topological superconductors and category theory},
  author={Bernevig, Andrei and Neupert, Titus},
  journal={Lecture Notes of the Les Houches Summer School: Topological Aspects of Condensed Matter Physics},
  publisher={Oxford University Press},
  pages={63--121},
  year={2017}
}

@article{KITAEV20032,
title = {Fault-tolerant quantum computation by anyons},
journal = {Annals of Physics},
volume = {303},
number = {1},
pages = {2-30},
year = {2003},
issn = {0003-4916},
doi = {https://doi.org/10.1016/S0003-4916(02)00018-0},
url = {https://www.sciencedirect.com/science/article/pii/S0003491602000180},
author = {A.Yu. Kitaev},
}

@article{KITAEV20062,
title = {Anyons in an exactly solved model and beyond},
journal = {Annals of Physics},
volume = {321},
number = {1},
pages = {2-111},
year = {2006},
note = {January Special Issue},
issn = {0003-4916},
doi = {https://doi.org/10.1016/j.aop.2005.10.005},
url = {https://www.sciencedirect.com/science/article/pii/S0003491605002381},
author = {Alexei Kitaev},
}

@article{Kellendonk01022009,
author = {Johannes Kellendonk and Serge Richard},
title = "{Weber–Schafheitlin-type integrals with exponent 1}",
journal = {Integral Transforms and Special Functions},
volume = {20},
number = {2},
pages = {147--153},
year = {2009},
publisher = {Taylor \& Francis},
doi = {10.1080/10652460802321485},
URL = {https://doi.org/10.1080/10652460802321485},
}

@misc{1004.5518,
Author = {Michał Wrochna},
Title = "{Weber-Schafheitlin integrals with arbitrary exponent}",
Year = {2010},
Eprint = {arXiv:1004.5518},
}

@article{PhysRevB.43.2661,
  title = "{Braid group and anyons on a cylinder}",
  author = {Hatsugai, Yasuhiro and Kohmoto, Mahito and Wu, Yong-Shi},
  journal = {Phys. Rev. B},
  volume = {43},
  issue = {4},
  pages = {2661--2677},
  numpages = {0},
  year = {1991},
  month = {Feb},
  publisher = {American Physical Society},
  doi = {10.1103/PhysRevB.43.2661},
  url = {https://link.aps.org/doi/10.1103/PhysRevB.43.2661}
}

@article{PhysRevB.43.10761,
  title = "{Anyons on a torus: Braid group, Aharonov-Bohm period, and numerical study}",
  author = {Hatsugai, Yasuhiro and Kohmoto, Mahito and Wu, Yong-Shi},
  journal = {Phys. Rev. B},
  volume = {43},
  issue = {13},
  pages = {10761--10768},
  numpages = {0},
  year = {1991},
  month = {May},
  publisher = {American Physical Society},
  doi = {10.1103/PhysRevB.43.10761},
  url = {https://link.aps.org/doi/10.1103/PhysRevB.43.10761}
}

@article{PhysRevB.107.195129,
  title = "{Numerical simulation of non-Abelian anyons}",
  author = {Kirchner, Nico and Millar, Darragh and Ayeni, Babatunde M. and Smith, Adam and Slingerland, Joost K. and Pollmann, Frank},
  journal = {Phys. Rev. B},
  volume = {107},
  issue = {19},
  pages = {195129},
  numpages = {33},
  year = {2023},
  month = {May},
  publisher = {American Physical Society},
  doi = {10.1103/PhysRevB.107.195129},
  url = {https://link.aps.org/doi/10.1103/PhysRevB.107.195129}
}

@article{BECKEN2000449,
title = "{The analytic continuation of the Gaussian hypergeometric function ${}_2F_1(a,b;c;z)$ for arbitrary parameters}",
journal = {Journal of Computational and Applied Mathematics},
volume = {126},
number = {1},
pages = {449-478},
year = {2000},
issn = {0377-0427},
doi = {https://doi.org/10.1016/S0377-0427(00)00267-3},
url = {https://www.sciencedirect.com/science/article/pii/S0377042700002673},
author = {W. Becken and P. Schmelcher},
}

@article{PhysRevLett.124.117205,
  title = "{Theory of Two-Dimensional Nonlinear Spectroscopy for the Kitaev Spin Liquid}",
  author = {Choi, Wonjune and Lee, Ki Hoon and Kim, Yong Baek},
  journal = {Phys. Rev. Lett.},
  volume = {124},
  issue = {11},
  pages = {117205},
  numpages = {5},
  year = {2020},
  month = {Mar},
  publisher = {American Physical Society},
  doi = {10.1103/PhysRevLett.124.117205},
  url = {https://link.aps.org/doi/10.1103/PhysRevLett.124.117205}
}

@article{PhysRevLett.122.257401,
  title = "{Resolving Continua of Fractional Excitations by Spinon Echo in THz 2D Coherent Spectroscopy}",
  author = {Wan, Yuan and Armitage, N. P.},
  journal = {Phys. Rev. Lett.},
  volume = {122},
  issue = {25},
  pages = {257401},
  numpages = {6},
  year = {2019},
  month = {Jun},
  publisher = {American Physical Society},
  doi = {10.1103/PhysRevLett.122.257401},
  url = {https://link.aps.org/doi/10.1103/PhysRevLett.122.257401}
}

@article{Mukamel,
   author = "Mukamel, Shaul",
   title = "Multidimensional Femtosecond Correlation Spectroscopies of Electronic and Vibrational Excitations", 
   journal= "Annual Review of Physical Chemistry",
   year = "2000",
   volume = "51",
   number = "Volume 51, 2000",
   pages = "691-729",
   doi = "https://doi.org/10.1146/annurev.physchem.51.1.691",
   url = "https://www.annualreviews.org/content/journals/10.1146/annurev.physchem.51.1.691",
   publisher = "Annual Reviews",
   issn = "1545-1593",
   type = "Journal Article",
  }

@Article{Kuehn2011,
author={Kuehn, W.
and Reimann, K.
and Woerner, M.
and Elsaesser, T.
and Hey, R.},
title="{Two-Dimensional Terahertz Correlation Spectra of Electronic Excitations in Semiconductor Quantum Wells}",
journal={The Journal of Physical Chemistry B},
year={2011},
month={May},
day={12},
publisher={American Chemical Society},
volume={115},
number={18},
pages={5448-5455},
issn={1520-6106},
doi={10.1021/jp1099046},
url={https://doi.org/10.1021/jp1099046}
}

@article{Woerner_2013,
doi = {10.1088/1367-2630/15/2/025039},
url = {https://doi.org/10.1088/1367-2630/15/2/025039},
year = {2013},
month = {feb},
publisher = {IOP Publishing},
volume = {15},
number = {2},
pages = {025039},
author = {Woerner, Michael and Kuehn, Wilhelm and Bowlan, Pamela and Reimann, Klaus and Elsaesser, Thomas},
title = {Ultrafast two-dimensional terahertz spectroscopy of elementary excitations in solids},
journal = {New Journal of Physics},
}

@article{Liu2025,
author={Liu, Albert},
title={Multidimensional terahertz probes of quantum materials},
journal={npj Quantum Materials},
year={2025},
month={Feb},
day={10},
volume={10},
number={1},
pages={18},
issn={2397-4648},
doi={10.1038/s41535-025-00741-y},
url={https://doi.org/10.1038/s41535-025-00741-y}
}

@article{Leinaas1977,
author={Leinaas, J. M.
and Myrheim, J.},
title={On the theory of identical particles},
journal={Il Nuovo Cimento B (1971-1996)},
year={1977},
month={Jan},
day={01},
volume={37},
number={1},
pages={1-23},
issn={1826-9877},
doi={10.1007/BF02727953},
url={https://doi.org/10.1007/BF02727953}
}

@article{PhysRevLett.48.1144,
  title = "{Magnetic Flux, Angular Momentum, and Statistics}",
  author = {Wilczek, Frank},
  journal = {Phys. Rev. Lett.},
  volume = {48},
  issue = {17},
  pages = {1144--1146},
  numpages = {0},
  year = {1982},
  month = {Apr},
  publisher = {American Physical Society},
  doi = {10.1103/PhysRevLett.48.1144},
  url = {https://link.aps.org/doi/10.1103/PhysRevLett.48.1144}
}

@article{PhysRevLett.49.957,
  title = "{Quantum Mechanics of Fractional-Spin Particles}",
  author = {Wilczek, Frank},
  journal = {Phys. Rev. Lett.},
  volume = {49},
  issue = {14},
  pages = {957--959},
  numpages = {0},
  year = {1982},
  month = {Oct},
  publisher = {American Physical Society},
  doi = {10.1103/PhysRevLett.49.957},
  url = {https://link.aps.org/doi/10.1103/PhysRevLett.49.957}
}

@article{Banerjee2017,
author={Banerjee, Mitali
and Heiblum, Moty
and Rosenblatt, Amir
and Oreg, Yuval
and Feldman, Dima E.
and Stern, Ady
and Umansky, Vladimir},
title={Observed quantization of anyonic heat flow},
journal={Nature},
year={2017},
month={May},
day={01},
volume={545},
number={7652},
pages={75-79},
issn={1476-4687},
doi={10.1038/nature22052},
url={https://doi.org/10.1038/nature22052}
}

@article{Banerjee2018,
author={Banerjee, Mitali
and Heiblum, Moty
and Umansky, Vladimir
and Feldman, Dima E.
and Oreg, Yuval
and Stern, Ady},
title="{Observation of half-integer thermal Hall conductance}",
journal={Nature},
year={2018},
month={Jul},
day={01},
volume={559},
number={7713},
pages={205-210},
issn={1476-4687},
doi={10.1038/s41586-018-0184-1},
url={https://doi.org/10.1038/s41586-018-0184-1}
}

@article{PhysRevB.55.2331,
  title = "{Two point-contact interferometer for quantum Hall systems}",
  author = {de C. Chamon, C. and Freed, D. E. and Kivelson, S. A. and Sondhi, S. L. and Wen, X. G.},
  journal = {Phys. Rev. B},
  volume = {55},
  issue = {4},
  pages = {2331--2343},
  numpages = {0},
  year = {1997},
  month = {Jan},
  publisher = {American Physical Society},
  doi = {10.1103/PhysRevB.55.2331},
  url = {https://link.aps.org/doi/10.1103/PhysRevB.55.2331}
}

@article{PhysRevX.13.011028,
  title = "{Interference Measurements of Non-Abelian $e/4$ \& Abelian $e/2$ Quasiparticle Braiding}",
  author = {Willett, R. L. and Shtengel, K. and Nayak, C. and Pfeiffer, L. N. and Chung, Y. J. and Peabody, M. L. and Baldwin, K. W. and West, K. W.},
  journal = {Phys. Rev. X},
  volume = {13},
  issue = {1},
  pages = {011028},
  numpages = {19},
  year = {2023},
  month = {Mar},
  publisher = {American Physical Society},
  doi = {10.1103/PhysRevX.13.011028},
  url = {https://link.aps.org/doi/10.1103/PhysRevX.13.011028}
}

@article{Werkmeister,
doi = {10.1126/science.adp5015},
author = {Thomas Werkmeister  and James R. Ehrets  and Marie E. Wesson  and Danial H. Najafabadi  and Kenji Watanabe  and Takashi Taniguchi  and Bertrand I. Halperin  and Amir Yacoby  and Philip Kim},
title = {Anyon braiding and telegraph noise in a graphene interferometer},
journal = {Science},
volume = {388},
number = {6748},
pages = {730-735},
year = {2025},
doi = {10.1126/science.adp5015},
URL = {https://www.science.org/doi/abs/10.1126/science.adp5015},
}

@misc{240319628,
Author = {Noah L. Samuelson and Liam A. Cohen and Will Wang and Simon Blanch and Takashi Taniguchi and Kenji Watanabe and Michael P. Zaletel and Andrea F. Young},
Title = "{Slow quasiparticle dynamics and anyonic statistics in a fractional quantum Hall Fabry-Pérot interferometer}",
Year = {2024},
Eprint = {arXiv:2403.19628},
}

@article{Nakamura2020,
author={Nakamura, J.
and Liang, S.
and Gardner, G. C.
and Manfra, M. J.},
title={Direct observation of anyonic braiding statistics},
journal={Nature Physics},
year={2020},
month={Sep},
day={01},
volume={16},
number={9},
pages={931-936},
issn={1745-2481},
doi={10.1038/s41567-020-1019-1},
url={https://doi.org/10.1038/s41567-020-1019-1}
}

@article{PhysRevB.74.045319,
  title = "{Electronic Mach-Zehnder interferometer as a tool to probe fractional statistics}",
  author = {Law, K. T. and Feldman, D. E. and Gefen, Yuval},
  journal = {Phys. Rev. B},
  volume = {74},
  issue = {4},
  pages = {045319},
  numpages = {16},
  year = {2006},
  month = {Jul},
  publisher = {American Physical Society},
  doi = {10.1103/PhysRevB.74.045319},
  url = {https://link.aps.org/doi/10.1103/PhysRevB.74.045319}
}

@article{PhysRevB.108.L241302,
  title = "{Anyonic Mach-Zehnder interferometer on a single edge of a two-dimensional electron gas}",
  author = {Batra, Navketan and Wei, Zezhu and Vishveshwara, Smitha and Feldman, D. E.},
  journal = {Phys. Rev. B},
  volume = {108},
  issue = {24},
  pages = {L241302},
  numpages = {6},
  year = {2023},
  month = {Dec},
  publisher = {American Physical Society},
  doi = {10.1103/PhysRevB.108.L241302},
  url = {https://link.aps.org/doi/10.1103/PhysRevB.108.L241302}
}

@article{BONDERSON20082709,
title = "{Interferometry of non-Abelian anyons}",
journal = {Annals of Physics},
volume = {323},
number = {11},
pages = {2709-2755},
year = {2008},
issn = {0003-4916},
doi = {https://doi.org/10.1016/j.aop.2008.01.012},
url = {https://www.sciencedirect.com/science/article/pii/S0003491608000171},
author = {Parsa Bonderson and Kirill Shtengel and J.K. Slingerland},
}

@misc{2505.01042,
Author = {Aprem P. Joy and Achim Rosch},
Title = "{Raman spectroscopy of anyons in generic Kitaev spin liquids}",
Year = {2025},
Eprint = {arXiv:2505.01042},
}

@article{PhysRevB.109.075108,
  title = {Anomalous thermal relaxation and pump-probe spectroscopy of two-dimensional topologically ordered systems},
  author = {McGinley, Max and Fava, Michele and Parameswaran, S. A.},
  journal = {Phys. Rev. B},
  volume = {109},
  issue = {7},
  pages = {075108},
  numpages = {25},
  year = {2024},
  month = {Feb},
  publisher = {American Physical Society},
  doi = {10.1103/PhysRevB.109.075108},
  url = {https://link.aps.org/doi/10.1103/PhysRevB.109.075108}
}

@misc{2503.22792,
Author = {Xu Yang and Ryan Buechele and Nandini Trivedi},
Title = {Detection of anyon braiding through pump-probe spectroscopy},
Year = {2025},
Eprint = {arXiv:2503.22792},
}

@article{PhysRevLett.116.156802,
  title = "{Current Correlations from a Mesoscopic Anyon Collider}",
  author = {Rosenow, Bernd and Levkivskyi, Ivan P. and Halperin, Bertrand I.},
  journal = {Phys. Rev. Lett.},
  volume = {116},
  issue = {15},
  pages = {156802},
  numpages = {5},
  year = {2016},
  month = {Apr},
  publisher = {American Physical Society},
  doi = {10.1103/PhysRevLett.116.156802},
  url = {https://link.aps.org/doi/10.1103/PhysRevLett.116.156802}
}

@article{Lee2022,
author={Lee, June-Young M.
and Sim, H.-S.},
title="{Non-Abelian anyon collider}",
journal={Nature Communications},
year={2022},
month={Nov},
day={04},
volume={13},
number={1},
pages={6660},
issn={2041-1723},
doi={10.1038/s41467-022-34329-y},
url={https://doi.org/10.1038/s41467-022-34329-y}
}

@article{Bartolomei,
doi = {10.1126/science.aaz5601},
author = {H. Bartolomei  and M. Kumar  and R. Bisognin  and A. Marguerite  and J.-M. Berroir  and E. Bocquillon  and B. Plaçais  and A. Cavanna  and Q. Dong  and U. Gennser  and Y. Jin  and G. Fève },
title = {Fractional statistics in anyon collisions},
journal = {Science},
volume = {368},
number = {6487},
pages = {173-177},
year = {2020},
doi = {10.1126/science.aaz5601},
URL = {https://www.science.org/doi/abs/10.1126/science.aaz5601},
}

@article{PhysRevX.13.011031,
  title = "{Comparing Fractional Quantum Hall Laughlin and Jain Topological Orders with the Anyon Collider}",
  author = {Ruelle, M. and Frigerio, E. and Berroir, J.-M. and Pla\ifmmode \mbox{\c{c}}\else \c{c}\fi{}ais, B. and Rech, J. and Cavanna, A. and Gennser, U. and Jin, Y. and F\`eve, G.},
  journal = {Phys. Rev. X},
  volume = {13},
  issue = {1},
  pages = {011031},
  numpages = {18},
  year = {2023},
  month = {Mar},
  publisher = {American Physical Society},
  doi = {10.1103/PhysRevX.13.011031},
  url = {https://link.aps.org/doi/10.1103/PhysRevX.13.011031}
}

@article{PhysRevX.13.011030,
  title = "{Cross-Correlation Investigation of Anyon Statistics in the $\ensuremath{\nu}=1/3$ and $2/5$ Fractional Quantum Hall States}",
  author = {Glidic, P. and Maillet, O. and Aassime, A. and Piquard, C. and Cavanna, A. and Gennser, U. and Jin, Y. and Anthore, A. and Pierre, F.},
  journal = {Phys. Rev. X},
  volume = {13},
  issue = {1},
  pages = {011030},
  numpages = {19},
  year = {2023},
  month = {Mar},
  publisher = {American Physical Society},
  doi = {10.1103/PhysRevX.13.011030},
  url = {https://link.aps.org/doi/10.1103/PhysRevX.13.011030}
}

\end{document}